\documentclass[12pt]{article}\usepackage[hyperfootnotes=false]{hyperref}
\usepackage{epsfig}
\usepackage{float}
\usepackage{empheq}
\usepackage{bbold}

\usepackage[utf8]{inputenc}
\usepackage{amsmath}

\usepackage{caption}

\usepackage{amsmath}
\usepackage{amssymb}
\usepackage{graphicx}
\newlength{\abstractwidth}
\renewcommand{\thefootnote}{\fnsymbol{footnote}}
\renewcommand{\thanks}[1]{\footnote{#1}} 
\newcommand{\starttext}{
\setcounter{footnote}{0}
\renewcommand{\thefootnote}{\arabic{footnote}}}

\newcommand{\be}{\begin{equation}}
\newcommand{\bea}{\begin{eqnarray}}
\newcommand{\eea}{\end{eqnarray}}
\newcommand{\beq}{\begin{equation}}
\newcommand{\ee}{\end{equation}}

\newcommand*\widefbox[1]{\fbox{\hspace{2em}#1\hspace{2em}}}

\def\dsp.{de Sitter space.}
\def\eq{&=&}

\def\la{\langle}
\def\ra{\rangle}
\def\simleq{\; \raise0.3ex\hbox{$<$\kern-0.75em
\raise-1.1ex\hbox{$\sim$}}\; }
\def\simgeq{\; \raise0.3ex\hbox{$>$\kern-0.75em
\raise-1.1ex\hbox{$\sim$}}\; }

\def\bi{\begin{itemize}}
\def\ei{\end{itemize}}
\def\S{Schwarzschild}
\def\sc{\setcounter{equation}{0}}
\def\dof{degrees of freedom }

\def\CJ{{\cal{J}}}

\def\CO{{\cal{O}}}

\def\bx{{\bar{\chi}}}

\def\bsub{ \begin{subequations}
\begin{empheq}[box=\widefbox]{align}  }
\def\esub{ \end{empheq}
\end{subequations}}

\def\1{\(  \mathbb{1} \)}

   \def\l{L_{c}}

 \def\lf{\left(}
    \def\rg{\right)}

  \def\bn{\bigskip \noindent}

    \def\dk{${\rm DSSYK_{\infty}}$}

\makeatletter
\g@addto@macro\normalsize{%
  \setlength\abovedisplayskip{10pt}
  \setlength\belowdisplayskip{20pt}
  \setlength\abovedisplayshortskip{10pt}
  \setlength\belowdisplayshortskip{20pt}
}
\makeatother

\usepackage{color}

\begin{document}


\begin{titlepage}


 \rightline{}
\bigskip
\bigskip\bigskip\bigskip\bigskip
\bigskip


\centerline{\Large \bf {De Sitter Space has no Chords. }} 

\bn

\centerline{\Large \bf {Almost Everything is Confined. }}

\bn



\bigskip
\begin{center}
	\bf   Leonard Susskind \rm

\bigskip

 Stanford Institute for Theoretical Physics and Department of Physics, \\
Stanford University,
Stanford, CA 94305-4060, USA \\ 

\bn

and Google, \\
Mountain View, CA

\end{center}

\bn


\begin{abstract}

This paper describes a phenomenon in which all but a tiny fraction of the fundamental holographic  degrees of  the SYK theory are confined (as in quark confinement) in the double-scaled infinite temperature limit. The mechanism for confinement is an essential ingredient in the duality between DSSYK and de Sitter space. 
The mechanism, which removes almost all states from the physical spectrum of the bulk de Sitter theory applies to configurations of a small number of fermions which would be expected to comprise Hawking radiation in de Sitter space. Without confinement there would be far too many species of Hawking particles. The mechanism   also applies to configurations with larger number of fermions, including the objects described by chord diagrams.

\end{abstract}

\end{titlepage}



 \rightline{}
\bigskip
\bigskip\bigskip\bigskip\bigskip
\bigskip





\bn


\bn

\starttext \baselineskip=17.63pt \setcounter{footnote}{0}

\tableofcontents

\Large


\section{Introduction}
The need to extend  the Holographic Principle from Anti de Sitter space to the cosmologically more relevant de Sitter space is self evident. But at  the moment very little is understood about de Sitter holography. It would be  helpful to have a concrete model to explore.
In this paper I will explain the conjecture  that the double-scaled limit of the SYK model at infinite temperature (\dk) is a holographic model of de Sitter space \cite{Susskind:2022bia}\cite{Rahman:2022jsf}.  

\subsection{De Sitter Space has no Chords}
\sc

The term ``chord" refers to contractions between double-scaled SYK operators, each consisting of a large number of fundamental fermion operators \cite{Berkooz:2018jqr}\cite{Lin:2022rbf}. There are two types of chords: the first, which I will continue to refer to as chords, has to do with insertions of the Hamiltonian. They are often called Hamiltonian chords but I will simply call them chords (think of the letter ``h" as standing for Hamiltonian). The second refers to contractions of any other operator of weight of order $q.$ These are often called ``matter chords." I'll call them ``cords" without the h. 

\begin{figure}[H]
\begin{center}
\includegraphics[scale=.4]{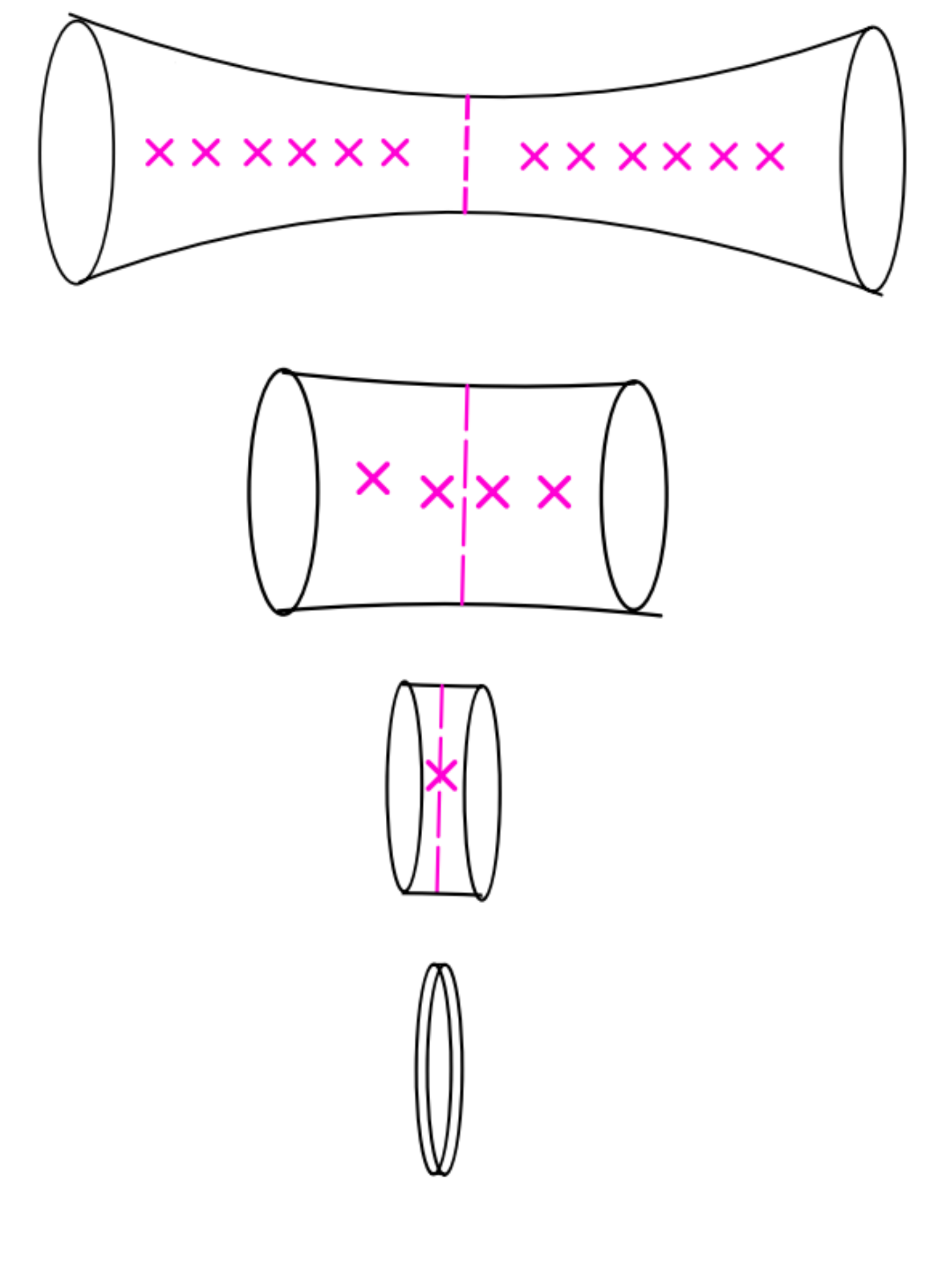}
\caption{Chords are represented by $X $. The number of chords is proportional to the length of the throat.}
\label{chords}
\end{center}
\end{figure}

Chords in DSSYK are thought to be related to the structure of space itself. Roughly they build space and the more chords there are the more space they describe. In particular they are instrumental in building the long wormhole or throat connecting the left and right side of a thermofield double. The average number of chords is proportional to the length of the wormhole which is equal to $\log{\beta}$. At very low temperature the wormhole is very long and the number of chords is large. As the temperature is increased the length decreases as does the number of chords (see figure \ref{chords}).

In the limit of infinite temperature the length of the  throat decreases to zero and the number of chords also tends to zero as in the last of figures \ref{chords}. One might think that there is no space left although the entropy remains large. What does that entropy describe holographically? My conjecture is that there is some remaining space and it is de Sitter space. Thus the slogan:

\bn
\it { `` de Sitter space has no chords." } \rm.
\bn 

The slogan is another way of saying that a static patch of de Sitter space is holographically described by a system at infinite temperature. The notation \dk \ means the double-scaled limit of SYK at infinite temperature.

\subsection{Almost Everthing is Confined}
I will assume that a static patch of de Sitter space is described by \dk \ with the \dof \  residing on the boundary of the static patch, i.e., the horizon. There are $N$ species of fermions which, at $T=\infty$  account for the large entropy $S=N.$
If these fermions could be emitted into the bulk (the static patch) they would comprise $N$ species of Hawking radiation. That is not consistent with a  local bulk description. In a conventional description of de Sitter space only the massless \dof---gravitons and photons---contribute to Hawking radiation, and the number of each  type of quanta in the bulk is order $1$. Almost all the femionic degrees of freedom must be confined near the horizon. Thus the slogan:

\bn
\it{Almost everything is confined.} \rm
\bn

 \section{Gravity Side of the Duality}
According to a recent conjecture the double-scaled infinite temperature limit of SYK theory (\dk) is dual to a particular version of de Sitter space; namely JT gravity with a 
positive cosmological 
constant \cite{Susskind:2022bia}\cite{Rahman:2022jsf}\footnote{A related conjecture was put forward by H. Verlinde---unpublished lecture.}. While the status of this conjecture is still not completely certain it may be worthwhile understanding the requirements for it to be correct. They may apply more broadly to other holographic de Sitter constructions.

 \subsection{JT from 3-D De Sitter}
 
 The static patch of dS(2+1) is described by the metric,
\be 
ds^2 = -(1-\frac{r^2}{\l^2})dt^2 +\frac{dr^2}{(1-\frac{r^2}{\l^2})} +r^2 d\alpha^2
\label{metric}
\ee
The parameter $\l$ is the cosmological length-scale, i.e., the radius of curvature of the de Sitter space.
The dimensional reduction is on the angular $\alpha$ coordinate, leading to a two-dimensional JT gravitational theory \cite{Rahman:2022jsf} \ with a dilaton field  $\Phi$  and positive cosmological constant.
 The solution of the JT equations of motion is two-dimensional de Sitter space with a dilaton equal to $r$,
\bea
ds^2 \eq  -(1-\frac{r^2}{\l^2})dt^2 +\frac{dr^2}{(1-\frac{r^2}{\l^2})} \cr \cr
\Phi &=& r.
\label{dilaton}
\eea

\subsection{Mass and Length Scales in dS}

Four mass scales  in de Sitter space will have important roles in what  follows. 
I will first describe them for the more familiar case of  4-dimensional de Sitter space and then return to 3-D.

\subsubsection{The Case of 4 Dimensions}
 The maximum mass in 4-D de Sitter space without severe global back-reaction is of order,
\be  
M_{max} \sim  \frac{\l}{G_4}   \ \ \ \ \ \ \
\label{Mmax4}
\ee
This nominally corresponds to the mass of a black hole with a \S \ radius comparable to the de Sitter radius $\l.$ It is also a completely classical scale in that no factor of $\hbar$ appears in \eqref{Mmax4}.

\begin{figure}[H]
\begin{center}
\includegraphics[scale=.45]{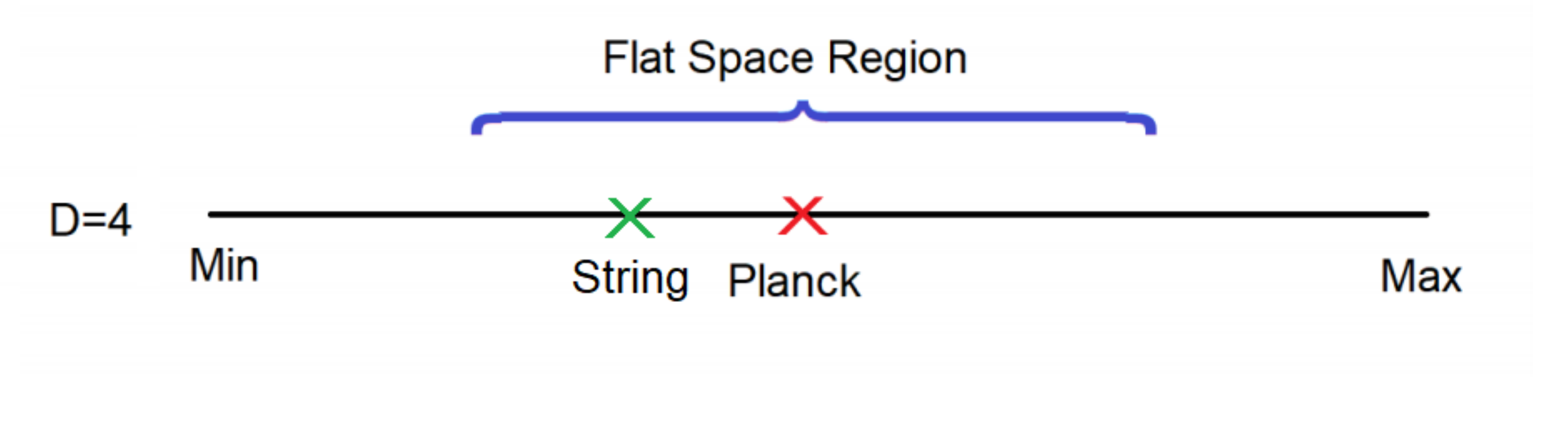}
\caption{Mass  scales in 4  dimensional dS. The horizontal axis represents $\log{M}.$ Mass scales increase to the right and length scales decrease. }
\label{scales4}
\end{center}
\end{figure}

The second limitation is at the opposite end of the energy spectrum: the energy should be large enough so that the corresponding wavelength is not longer than  $\l.$ The minimum energy that satisfies this criterion is
\be
M_{min} = \hbar /\l. 
\label{Mmin4}
\ee 
At these extreme ends of the spectrum the curvature of dS cannot be ignored, but for phenomena in the range
$$M_{min}<<M<<M_{max}$$ flat space is a good approximation. I'll call this ``the flat space region." It is centered around the geometric mean of $M_{min}$ and $M_{max}.$ which  I'll denote by $M_{m}$ ($m$ meaning microscopic, mean, or middle),
\bea
M_{m} &=& \sqrt{M_{min}M_{max}}  \cr \cr
\eq \sqrt{\frac{\hbar}{G_4}}
\label{=1/G4}
\eea
This is precisely the four-dimensional  Planck mass,
\be 
M_m = M_{planck}   \ \ \ \ \ \ \ \ \ \ (\rm in \ 4 \ dimensions)
\label{Mm=Mp}
\ee
It lies at the center (logarithmically) of the flat-space region, midway between $M_{min}$ and $M_{max}$ as shown in figure \ref{scales4}
The two length scales, quantum-mechanically  related to $M_{min} \ \text{and} \  M_{m}$ 
are,
\bea 
\l \eq \hbar /M_{min} \cr \cr
 L_{m} \eq  L_{planck}   \cr 
 \eq \hbar /M_{planck} \cr
 \eq \hbar /M_{m}
\label{length4}
\eea

In 4-dimensions the Planck mass plays a dual role. On the one hand (as always) $L_{planck}$ controls the density of entropy on horizons through the Bekenstein law. On the other hand $M_{planck}$ defines the mid-point $M_{m}$ lying between the extremes $M_{max}$ and $M_{min}$. These are two  different concepts, but in 4-D they just  happen to define the same  mass and length scales.

It is not clear that lengths smaller than $L_{planck}$ are meaningful. In particular $\hbar /M_{max}$ does not seem to play any interesting role. The natural length scale to associate with $M_{max} $ is the purely classical scale $G_4 M_{max} .$

\bn

Another scale that will \it will \rm \ concern us is the string scale $L_s.$ In 4- dimensions it is given in terms of the  string-coupling and the Planck scale,
\bea
M_s \eq    \frac{\hbar}{L_s} \cr \cr
 \eq g M_{planck}.
\label{lstring}
\eea
It is common to assume that $g$ is finite in the semiclassical limit which implies that the ratio of the string and micro  scales are fixed.

The various length scales, $\l, \ L_m, \ L_s$ allow us to define several systems of units. For example time measured in units of $\l$ will be called cosmic time, $t_c$. Similar considerations apply to micro and string time:
\bea
t_c \eq  \lf  \frac{t}{\l} \rg \l    \cr \cr
t_m \eq  \lf  \frac{t}{L_m} \rg  \l   \cr \cr
t_s \eq  \lf  \frac{t}{L_s} \rg  \l
\label{tunits}
\eea
The quantities in the parenthesis represent time measured in cosmic units,  micro units, and  string units. The universal factor of $L_c$ is just for dimensional consistency, in order to give  $t_c, \ t_m,$ and $t_s$  dimensions of time.

(Note that while $\l, \ L_m, \ L_s$ represent definite fixed lengths, the quantities $t_c, \ t_m, \ t_s$ represent  variable time  measured in different units.)

\bn
\bf Note: 
Henceforth  we will set $\hbar =1.$

\rm

\bn

\subsubsection{3-D} \label{3-D}
\begin{figure}[H]
\begin{center}
\includegraphics[scale=.6]{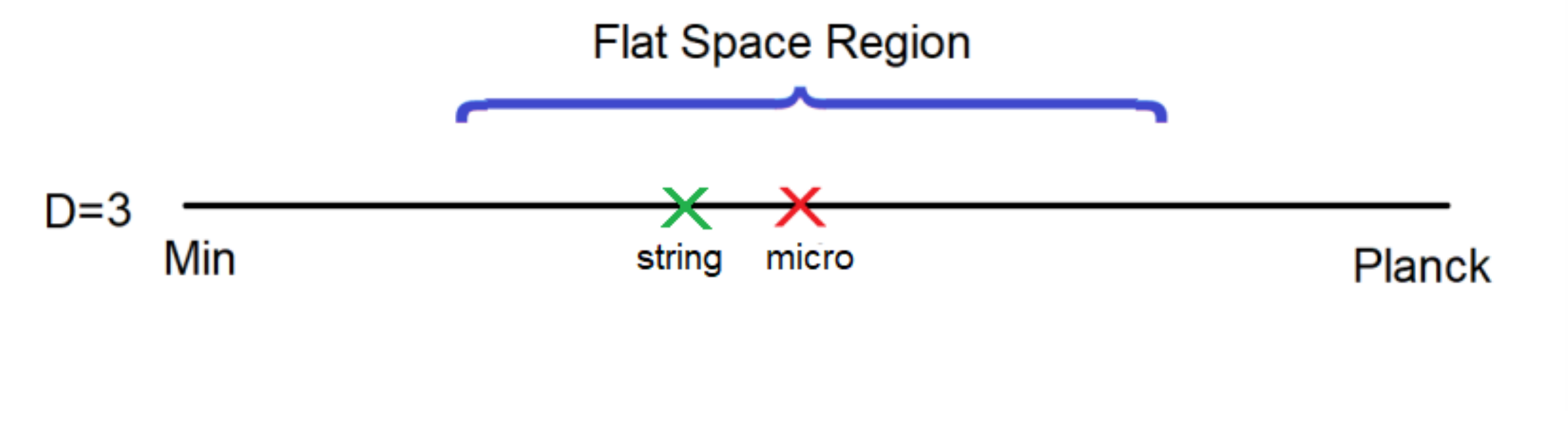}
\caption{Mass  scales in 3  dimensional including the string scale. Typically the string mass scale will lie near but somewhat below the geometric mean of $M_{min}$ and $M_{max}.$} 
\label{allscales}
\end{center}
\end{figure}

From now on we will be concerned only with 3-dimensions where
the situation shown in figure \ref{allscales}  looks similar to, but different  from figure \ref{scales4}.

\begin{figure}[H]
\begin{center}
\includegraphics[scale=.25]{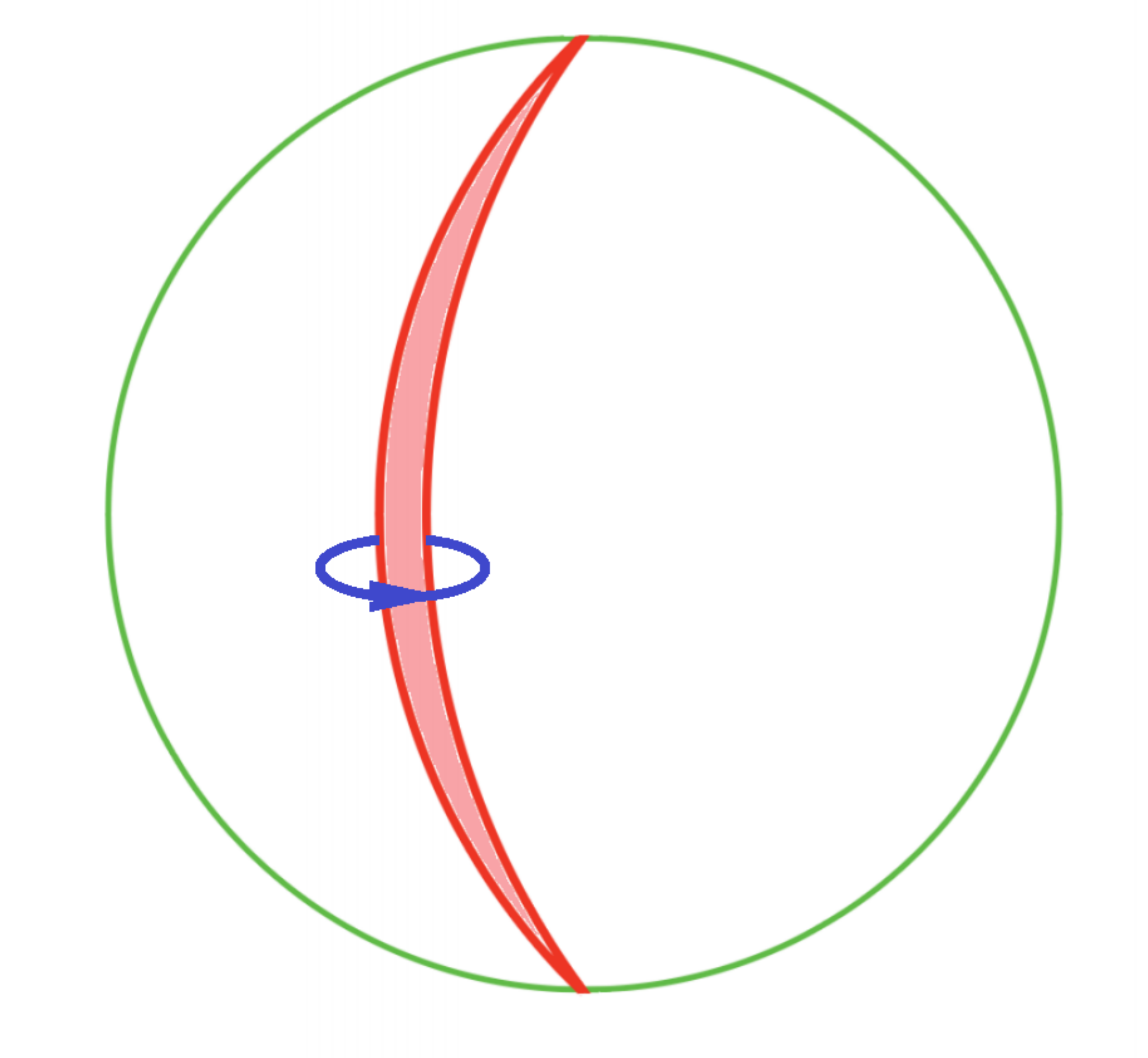}
\caption{A mass close to the Planck mass in 3-dimensional de Sitter space creates a conical deficit close to $2\pi.$ The geometry is a thin sliver with the edges identified.}
\label{sliver}
\end{center}
\end{figure}

\bn
Figure \ref{allscales} indicates that  the largest mass which can be placed in 3-D de Sitter is    $1/G_3)$ which is technically equal to the three dimensional Planck mass. 
\be 
M_{max} =  \frac{1}{G_3}.
\label{mmax3}
\ee
The reason is that in three dimensions  a mass creates a conical deficit---the  mass $M_{max}$  giving a limiting deficit of $2\pi.$ The back-reaction on the geometry is so strong that the geometry becomes a periodically-identified thin sliver when the mass approaches $ \frac{1}{G_3}$ as shown in figure \ref{sliver}.
As before, the smallest mass for which the wavelength is  $\geq \l$ is,
\be
M_{min} = 1/\l. 
\label{mmin3}
\ee
 The micro-mass---the geometric mean of $M_{max}$ and $M_{min}$---is,
\bea 
M_{m}&=&   1/L_{m}    \cr \cr
 &=&  \frac{1}{\sqrt{\l}}  \frac{1}{\sqrt{G_3}} .
\label{mm}
\eea
In 3-D the mass $M_m$ is not the Planck mass $ 1/G_3.$

\bn

\bn

In passing from 4-D to 3-D the Planck mass generalizes in two distinctly  different ways: 
\begin{enumerate}
\item  The usual Planck mass governing the density of horizon entropy.
\item  The micro mass $M_m$ lying midway between $M_{max}$ and $M_{min}$.
\end{enumerate}

\bn \bn

In $D=3$ equations \eqref{length4} are replaced by,
\bea 
\l \eq 1/M_{min}  \cr \cr
L_{m} \eq 1/M_{m} = \sqrt{\l} \sqrt{G_3}
\label{length3}
\eea

As in 4-D we can define cosmological and microscopic units of time,
\bea
t_c \eq \lf  \frac{t}{\l} \rg  \l   = t  \cr \cr
t_m \eq \lf \frac{t}{L_{m}} \rg   \l
\label{units3}
\eea
Their ratio satisfies,
\bea
\frac{t_m}{t_c} \eq \sqrt{\frac{\l}{G_3}} \cr \cr
&\sim& \sqrt{S_{ds}}.
\label{ratiot3}
\eea
where  $S_{ds}$ is the usual de Sitter entropy.
The ratio of microscopic and cosmic scales diverges in the semiclassical limit $S_{ds} \to \infty.$

\subsection{String Scale}\label{Ss :string}

\dk \ is not in any obvious way a string theory (non-obvious is quite  another matter). Nonetheless  there is 
an additional mass scale (and related length scale) in \dk \ which  plays a role similar to  that of  the string scale in four dimensions.  I'll call it  $M_s$ and the corresponding length scale $L_s$  ($M_s = 1/L_s$).
In \dk \  the scales $M_s$ and  $L_s$  are   dynamical, emerging from SYK dynamics similar to the  way the confinement scale emerges from gauge theory dynamics.  The origin of the string scale is the subject of section \ref{S:strscale}.

The string length-scale $L_s$ is assumed to be microscopic; in other words proportional to $L_m,$ although typically somewhat larger as shown in figure \ref{allscales}.
We define a parameter $\lambda$ by,
\be 
\lambda = \lf \frac{L_m}{L_s} \rg^2 = \lf  \frac{M_s}{M_m}   \rg^2
\label{ls3}
\ee
Our assumption will be that the ratio $L_m/L_s$ (and therefore $\lambda$) is  finite in the semiclassical limit.

 We will see that
$\lambda$ also has a dual meaning as a parameter in \dk.
If  $\lambda$ is held fixed at order $1$ then $L_s$ will be of the same order as the microscopic scale $L_{m}$. In this case the theory will  be local in the sense that the string scale will be vanishingly small on the cosmic scale. To put it another way the theory exhibits ``sub-dS locality" in the semiclassical limit.

On the other hand $\lambda$ is a tunable parameter which if taken  very small will lead to a string scale much larger than $L_m.$  As $\lambda \to 0$ the string scale can approach the cosmic scale, the theory becoming highly non-local in the bulk.

\bn

To summarize, there are four important mass scales that we will be concerned with. In increasing order they are:
\begin{enumerate}
\item The minimum mass scale $M_{min},$ equal to the de Sitter-Hawking temperature. 
\item The string scale $M_s$. This is the scale involving objects composed of $q$ elementary quanta. It is the subject of work on  ``chords" and chord diagrams \cite{Berkooz:2018jqr}\cite{Lin:2022rbf}. 
\item The middle/mean/micro scale $M_m.$
\item The maximum scale $M_{max}$, also equal to the three-dimensional Planck scale.
\end{enumerate}

Equivalently we may use $M_{min}$, $M_s$, $M_{max},$ and the dimensionless ratio $\lambda = \lf \frac{L_m}{L_s} \rg^2$ to parameterize  the scales of interest.

\subsection{The Separation of Scales}
In the semiclassical limit (the limit of infinite entropy) a separation of scales takes place. 
The ratios  $\frac{M_{max}}{M_m}$ and   $\frac{M_m}{M_{min}}$   diverge,
\bea 
 \frac{M_{max} }{M_m} &\to& \infty \cr \cr
  \frac{M_{m} }{M_{min}} &\to& \infty
  \label{SCLratios}
\eea
The ratio of the string scale to the micro scale is assumed to be finite in the semiclassical limit,
\be 
\frac{M_s}{M_m} \to \rm finite
\label{finiterat}
\ee

The dynamical behaviors of the three scales $M_{max}, M_{min}, M_m$ are very different and in a certain sense they decouple in the semiclassical limit. Most of the recent interest in the double-scaled limit \cite{Berkooz:2018jqr}\cite{Lin:2022rbf}
is concerned with the string scale, but 
in this paper I will be equally interested in physics at the cosmic scale $M_{min} = 1/L_c,$  i.e., the Hawking temperature. This includes quasi normal modes and Hawking radiation.

\subsection{Hawking Radiation}
De Sitter Hawking radiation is radiation emitted into the static patch by the horizon of de Sitter space. The characteristic wavelength is $\l.$ The radiation  is extremely sparse; roughly one quantum in the observable universe for each massless degree of  freedom. In other words the observable universe contains about one or two Hawking  photons and a similar number of gravitons. If there happen to be massless fermions they will also show up.

If \dk \ is the holographic description of some de Sitter space as I have conjectured, then there are a huge number of perturbatively massless fundamental fermionic degrees of freedom. If they all manifested themselves as light bulk fields which  propagate into the static patch, there would be a runaway situation in which the static patch would be occupied by an enormous number of low energy quanta. This is not consistent with bulk locality\footnote{I am grateful to Edward Witten for a helpful discussion concerning this point.}. The main point of this paper is that in the \dk \ limit this does not occur---instead almost all the fermionic \dof \ are dynamically confined to the stretched horizon.

In fact there may not be any particles light enough to be emitted into the bulk as Hawking radiation. For example in $(2+1)$-dimensional de Sitter space there are no gravitons, and in the simplest version without a $U(1)$ symmetry  there are no photons.  Hence we may expect that despite the huge number of boundary fields, a holographic theory based on \dk \ with real (as opposed to complex) fermions will have no Hawking radiation at all. The slightly richer ``complex" SYK  theory with a $U(1)$ symmetry might have a photon, but that is all.
This leads to the question: What happens to the large number of perturbatively massless fermion fields? 

In section \ref{plasma} I will describe an analogous situation that takes place in large $N$ QCD where one might expect a bubble of QCD plasma to evaporate almost instantly due to the large number of gluon fields. In both cases the answer is the same---almost all the degrees of freedom are confined.

\sc
\section{DSSYK$_{\bf {\infty}}$} \label{dssyk}

The  real (or Majorana) DSSYK model (see \cite{Maldacena:2016hyu} for details) is defined as a system of $N$ real fermionic degrees of freedom $\chi_i,$ 
\be 
\{ \chi_i,  \chi_j    \} = 2\delta_{ij}
\label{anticoms}
\ee
 coupled through $q$-local interactions,
\be 
H = \sum J_{i_1i_2 i_3...i_q}\chi_{i_1}\chi_{i_2}\chi_{i_3}...\chi_{i_q}.
\label{Hsyk}
\ee
The couplings $J$ are drawn from a gaussian ensemble with variance,
\be 
\la JJ\ra =\frac{q!}{N^{q-1}}\CJ^2.
\label{varMS}
\ee
where $\CJ$ is a fixed parameter with units of energy.

Equation \eqref{varMS}  differs from the standard scaling by a factor of $q^2.$ The standard formula for the variance, for example as given in \cite{Maldacena:2016hyu}, is
\be 
\la JJ\ra_{standard} =\frac{q!}{q^2N^{q-1}}\CJ^2.
\label{varstandard}
\ee
In the next subsection I will explain the apparent discrepancy between \eqref{varMS} and \eqref{varstandard}.

\bn

The \dk \ model is defined   \cite{Cotler:2016fpe}\cite{Berkooz:2018jqr}\cite{Lin:2022rbf} by letting $N\to \infty$ while allowing $q$ to  grow with $N$ according to,
\be 
\frac{q^2}{N} = \lambda \ \ \ \ \ \ \ \ \ ( \lambda \ \rm fixed)
\label{q2/N=lamb}
\ee

The symbol $\lambda$ has already been used earlier in \eqref{ls3} to represent the ratio ${L_m^2}/{L_s^2}$ whereas in \eqref{q2/N=lamb} it represents $q^2/N.$
We will see in section \ref{S:strscale} that the two definitions of $\lambda$ define the same quantity.

\subsection{The Hamiltonian}
It is important at this point to recall the separation of scales that takes place in the semiclassical limit. When one discusses energy, time, or the Hamiltonian 
it is necessary to specify the units in which time is measured.  

The standard Hamiltonian with variance given by \eqref{varstandard} represents evolution in string units. The Hamiltonian with variance \eqref{varMS}  represents evolution in cosmic units. One may think that this is a trivial difference since the two scales are related by a numerical factor $q$ but this would miss one of the main points that I have emphasized in the past. Different things have finite limits in different units. For example in four dimensions the oscillation frequency of strings has a finite limit in string units, but in cosmic units the frequency goes to infinity. On the other hand the decay rate of quasinormal modes is finite in cosmic units but goes to zero in string units.

The  Hamiltonian in \eqref{Hsyk} is scaled in such a way that it generates time-translations in cosmic  units and will produce finite limits for quasinormal modes.
\be 
H= i \frac{d}{dt_c} = i \frac{d}{dt}.
\label{H_c}
\ee
The standard scaling (for example \cite{Maldacena:2016hyu}) corresponds to string units and will give infinity for the lifetime of quasinormal modes.

\bn

Finally, for reasons explained in \cite{Lin:2022rbf}, to define \dk \  the temperature defined through the Boltzmann distribution,
\be 
\rho = \frac{1}{Z}e^{H/T}
\label{rho=boltz}
\ee
is taken to infinity,
\be 
T= \infty.
\label{T=inf}
\ee
As a consequence the density matrix of the static patch is maximally mixed as advocated in  \cite{Dong:2018cuv}\cite{Chandrasekaran:2022cip}\cite{Lin:2022nss}. One should not confuse the formal temperature in \eqref{rho=boltz} with the Hawking temperature $T_H$ or ``tomperature" as explained in \cite{Lin:2022nss}.

\bn

Most likely the model based on real SYK has no bulk fields light enough to be emitted as Hawking radiation. Being based on a dimensional reduction from $(2+1)$
dimensions there are no gravitons and since there is no Maxwell sector there are no photons. This is a bit disappointing since we would like to use Hawking radiation to probe the geometry of the static patch. However we can introduce Maxwell degrees of freedom by generalizing the  degrees of freedom from real to complex fermions, $\chi_i$ and $\bx _i$.

\subsection{Complex SYK}

The complex version of \dk \ is defined by  \cite{Berkooz:2020uly},

\bea
\{  \chi^i , \bx _j   \} &=& 2\delta^i_j \cr \cr
\{  \chi^i , \chi^j   \} &=& 0  \cr \cr
\{ \bx_i ,  \bx_j  \} \eq 0
\label{Canticom}
\eea

\be 
H = \sum J^{i_1 \cdot \cdot \cdot i_q}_{j_1 \cdot \cdot \cdot j_q}   \ \bx_{i_q}
\cdot \cdot \cdot \bx_{i_1}   \  \chi^{j_1}  \cdot \cdot \cdot  \chi^{j_q}
\label{Hcomp}
\ee
with $1\leq i_1 \cdots \leq i_q$, and  $1\leq j_1 \cdots \leq j_q$.

The couplings $J$ satisfy the ensemble averages,
\be 
\la  \lf J^{i_1 \cdot \cdot \cdot i_q}_{j_1 \cdot \cdot \cdot j_q} \rg
     \lf J^{i_1 \cdot \cdot \cdot i_q}_{j_1 \cdot \cdot \cdot j_q} \rg^*
     \ra
     =\CJ^2 N \lf \frac{q!}{N^{q}}  \rg^2  \ \ \ \ \ \ \text{(no sum)}
     \label{varcomp}
\ee
The time and mass scales for complex \dk  are essentially the same as for real \dk. 

\bn

The SYK model with real fermions has no continuous symmetry, at least  in the usual sense of the term symmetry. But the disordered ensemble average defined by \eqref{varMS} has $SO(N)$ symmetry.  Symmetries of disordered averages do not apply to individual instances of the ensemble and normally do not have the dynamical consequences of conventional symmetries such as conservation laws.

\bn

The symmetry of  complex SYK is larger than the real case. First of all it has the conventional global symmetry $U(1).$  In the bulk theory it implies a $U(1)$ gauge symmetry and electric forces between bulk charged particles  \cite{Susskind:2020fiu}.
The ensemble average symmetry is also larger in the complex case. Instead of $
SO(N)$ it is $SU(N).$

\bn

Following    \cite{Berkooz:2018jqr}  and  \cite{Lin:2022rbf}  we will sometimes consider cord operators similar to the Hamiltonian in \eqref{Hsyk},
\bea 
O \eq  \sum o_{i_1i_2 i_3...i_q}\chi_{i_1}\chi_{i_2}\chi_{i_3}...\chi_{i_q}
\cr \cr
\la oo\ra &=& \frac{Nq!}{N^{q}}\CO^2 \cr \cr
\eq
 \frac{q!}{N^{q-1}}\CO^2.
\label{O}
\eea
Such operators, formed from monomials of order $q$  have energy in the string range $q\CJ.$ 

\subsection{Dictionary}\label{dictionary}
The dictionary relating gravitational bulk parameters to \dk \ parameters is simple and  was worked out in \cite{Susskind:2022bia} (see section 6). I will just quote the results here.

The minimum scale $M_{min},$ which also equals the Hawking temperature is given by,
\be 
M_{min} = \CJ.
\label{Mmn=J}
\ee
This can be seen a number of ways. By studying the properties of scrambling in \dk \ it was shown  in \cite{Susskind:2022dfz} that the decay rate of quasinormal modes  is of order $\CJ.$

On the gravitational side $M_{max}$  is given by the condition that the conical deficit should be $2\pi.$ This gives $M_{max} = 1/G.$ It follows that the ratio $M_{max}/M_{min}$ is 
\be 
\frac{M_{max}}{M_{min}} = \frac{L_c}{G}
\label{max/min}
\ee
which apart from a numerical factor is the de Sitter entropy. Thus 
$M_{max}/M_{min}$ is 
\be 
\frac{M_{max}}{M_{min}} =N
\label{max/min=N}
\ee
and
\be 
M_{max} = \CJ N
\label{Mmx=JN}
\ee

The micro or mean scale $M_m$ is the geometric mean of $M_{max}$ and $M_{min}.$
\be 
M_m = \CJ \sqrt{N}
\label{Mm=JsqrtN}
\ee

Later we will see that the string scale $M_s$ is given by,
\be 
M_s = q\CJ.
\label{Ms1}
\ee









\section{Perturbation Theory}
In this section I will review some facts about    SYK perturbation theory at infinite temperature \cite{Roberts:2018mnp} and in cosmic units. The units are important. For example in string units each diagram would have an additional factor of $1/q$ for each vertex. The structure of the perturbation series that I will describe---in particular it's relation to large $N_{ym}$ perturbation theory---applies in cosmic units.

\begin{figure}[H]
\begin{center}
\includegraphics[scale=.4]{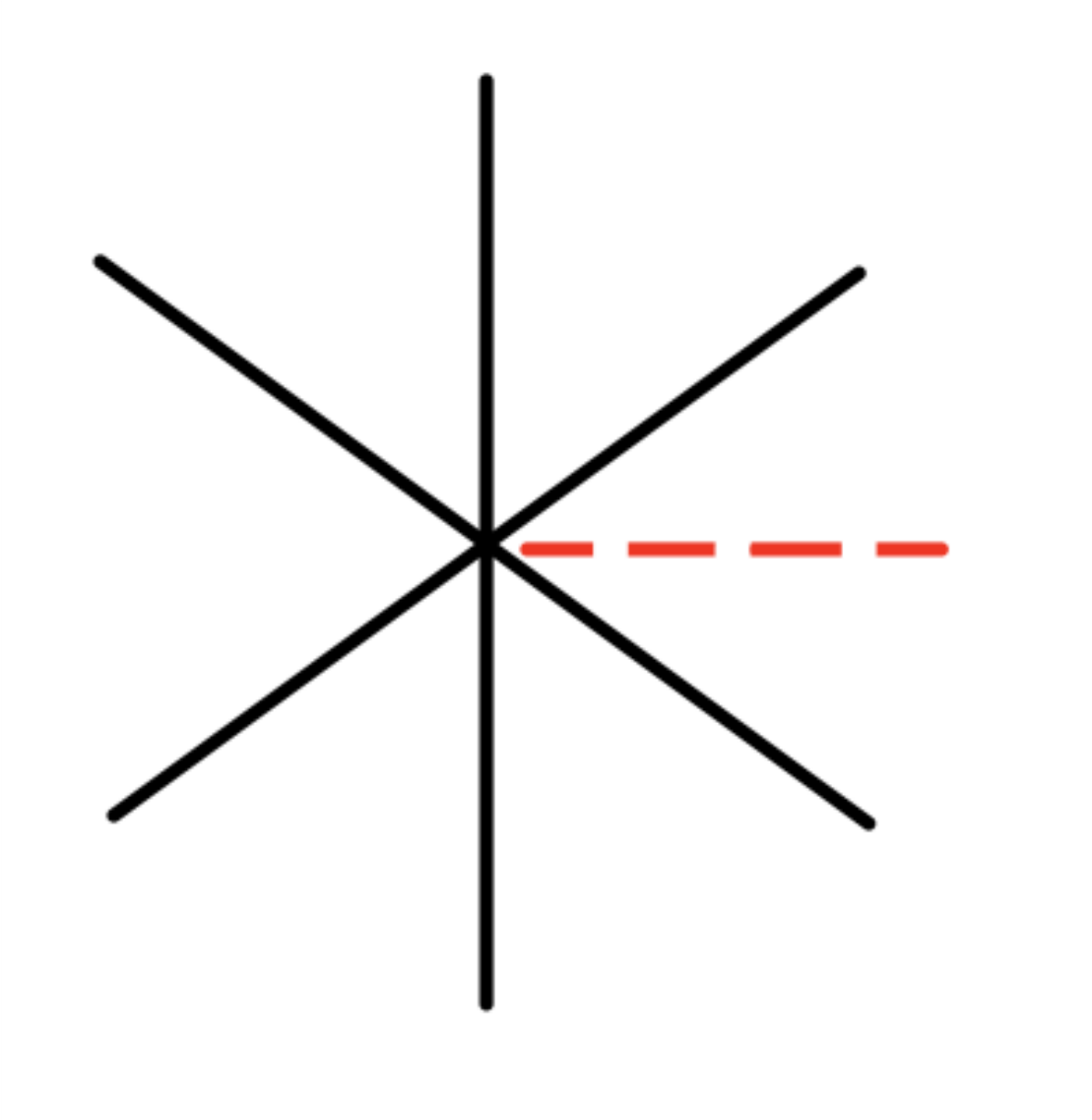}
\caption{A vertex for the case $q=6.$}
\label{star}
\end{center}
\end{figure}

The expansion parameter is $\CJ.$ 
At each vertex (figure \ref{star}) $q$ fermions are emitted. 
The fermion lines are shown in black. In addition each vertex emits a dashed red
 line which represents one end of  a $JJ$ correlation function \eqref{varMS}.

The fermionic propagators are represented by solid black lines. At infinite temperature they are very simple, given by  
\be 
\epsilon(t_2-t_1) = \text{sign}(t_2-t_1)
\label{fermprops}
\ee
 where $t$  refers to cosmic time.
Ensemble averages over the gaussian probabilities for 
$J$ are represented by dashed red lines.

\begin{figure}[H]
\begin{center}
\includegraphics[scale=.3]{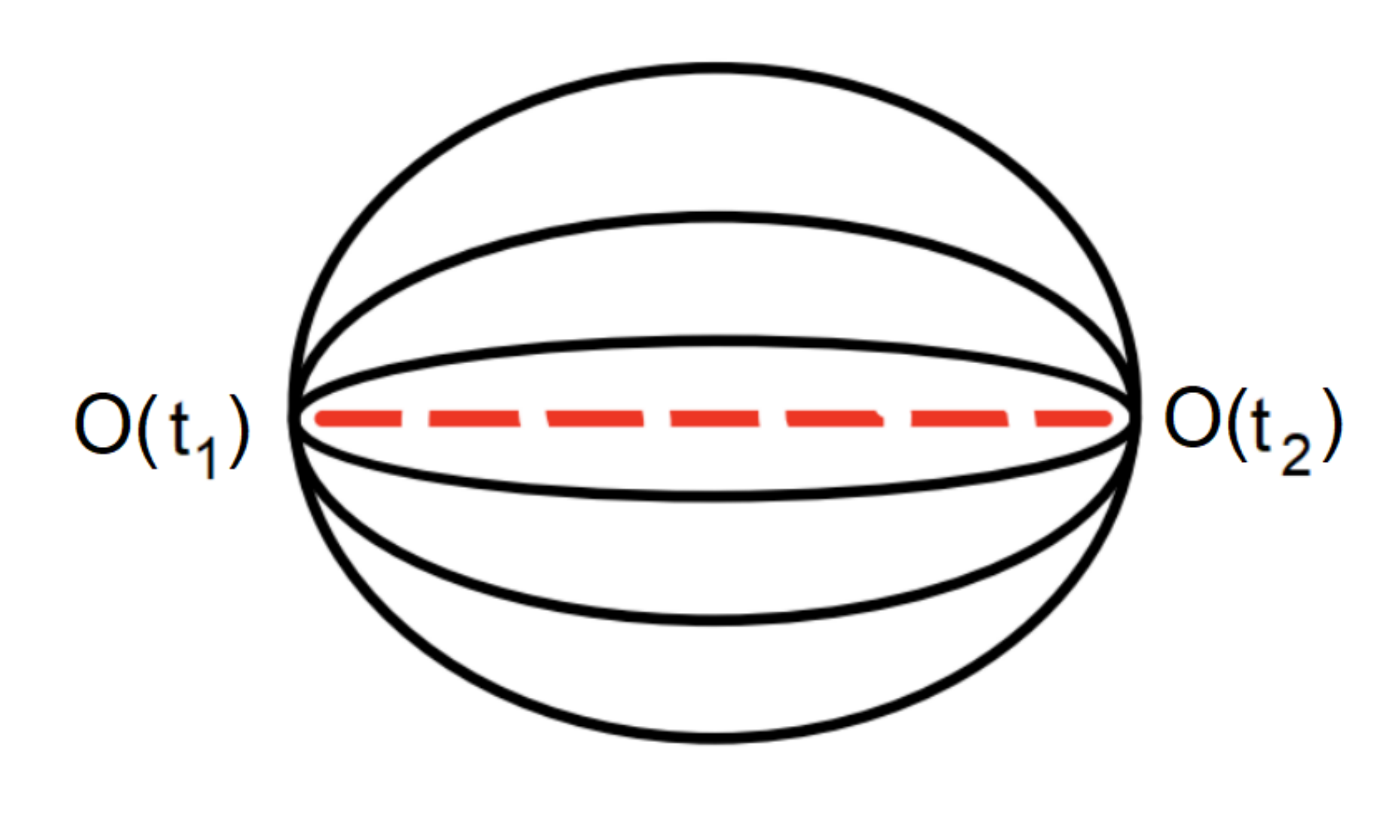}
\caption{The simplest vacuum melon diagram.}
\label{melon}
\end{center}
\end{figure}

The discussion will be very brief and will focus on a few specific diagrams as examples.
To illustrate let's consider the  diagram in figure \ref{melon} representing the correlation function $$T\la O(t_1) O(t_2)  \ra$$ (where $T$ means time-ordered product) in real SYK.
\bn
For definiteness I've drawn the diagram for the case $q=6$ but the generalization is obvious.
Since there are an even number of propagators  in figure \ref{melon}   the overall sign is positive. 

For real SYK the numerical  coefficient for  the diagram is
\be 
\text{diagram} = \CO^2   \     \frac{N^q}{q!}   \times    \frac{q!}{N^{q-1}}
\label{vacdiag}
\ee
The  factor  $\frac{N^q}{q!}$ is   the number of ways of choosing $q$ fermion modes from a total of $N$,
$$\frac{N!}{q!(N-q)!} \approx \frac{N^q}{q!}.$$
The second factor, corresponding to the dashed red line,  is 
 the correlator  $\la oo\ra$  in \eqref{O}.  
  The  result is 
\be
\text{diagram } = \CO^2  N.
\label{vacdiag2}
\ee
If we want  to integrate the diagram over the relative time between vertices the expression would be,
\be
\text{vac-diag } = \CO^2  N \int dt,
\label{vacdiag2}
\ee
The result is infrared divergent because the integrand is independent of $t$.

\begin{figure}[H]
\begin{center}
\includegraphics[scale=.4]{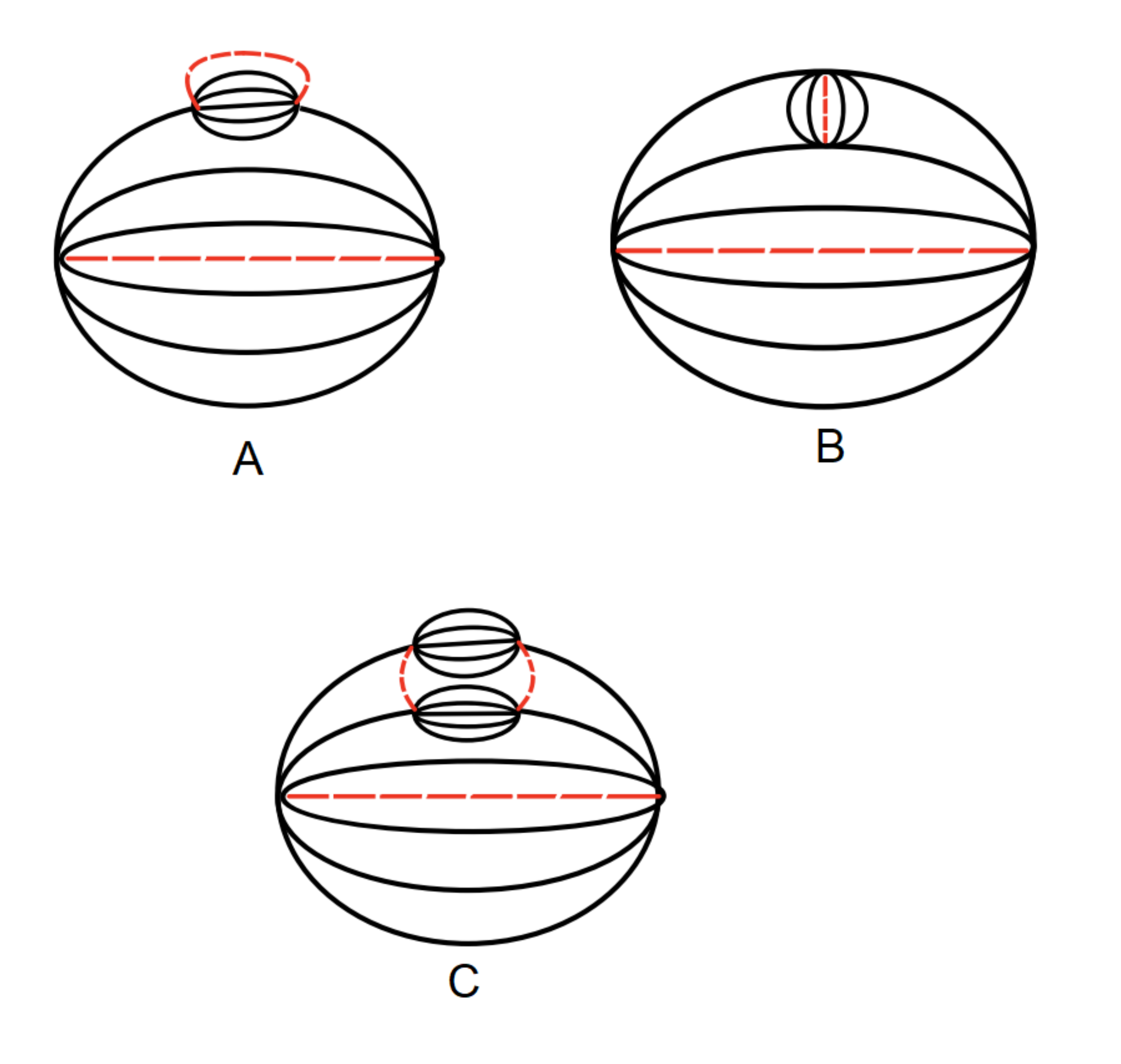}
\caption{Additional vacuum diagrams.  \ (A) another melon diagram. \  (B) a non-melonic diagram to next order in $1/N$. \ (C) a non-perturbative diagram in the $1/N$ expansion.}
\label{diagrams}
\end{center}
\end{figure}

In figure \ref{diagrams} three more typical diagrams are shown.
 Diagram 5$\bf A$ is another melon diagram whose leading behavior for complex SYK is,
\be
\CO^2 \CJ^2 q \frac{N^q}{q!} \frac{N^{q-1}}{(q-1)!} \times  \frac{q!}{N^{q-1}}  \frac{q!}{N^{q-1}} \int d^3 t,
\label{A}
\ee
where the integral is an appropriate three dimensional integral.  I will not spell it out further except to say that it is IR-divergent. All together we get,
\be 
\CO^2 \CJ^2 N q^2 \int d^3 t
\label{vacdiag3}
\ee
The expression \eqref{vacdiag3} scales the same way with $N $ as \ref{vacdiag2} but contains an extra factor of $ q^2$. 

The next example,  diagram   \ref{diagrams} $\bf{B},$  is non-melonic. It has the value,
\bea 
&&\CO^2 \CJ^2 q(q-1) \frac{N^q}{q!}     \frac{N^{q-2}}{(q-2)!} \times   \frac{q!}{N^{q-1}}     \frac{q!}{N^{q-1}}   \int d^3 t  \cr \cr
\eq \CO^2 \CJ^2 q^2(q-1)^2   \int d^3 t 
\label{vac4}
\eea
This non-melonic diagram is smaller by a factor of $N,$  but  higher order in $q$ than the leading melon diagram. 

Next let us look at figure \ref{diagrams}$\bf{C}$. This diagram is different than the previous ones in that the dotted red lines---the $jj$ correlators---are not contained within melonic structures. This is a signal that the diagram is non-perturbative in the $1/N$ expansion. 
It is given by,
\be  
\CO^2 \CJ^4   (q-1)q\frac{N^q}{q!}    \frac{N^{q-2}}{(q-2)!}  \times   \frac{q!}{N^{q-1}}    \frac{q!}{N^{q-1}}        \frac{q!}{N^{q-1}}   \int d^5 t  
\ee
Notice that there are more factors of $N^q$ in the denominator than the numerator. To leading order in $q$ the final result is,
\be
\sim  \CO^2  \CJ^4 \lambda^2 N^4  \lf  \sqrt{\frac{\lambda}{N}}   \rg^{\sqrt{\lambda N}}
 \int d^5 t 
\ee
This nonperturbative contribution (NPC) vanishes exponentially in the double-scaled limit. It is similar to instanton contributions to the large $N_{ym}$ expansion of gauge theory amplitudes.

\subsection{DSSYK$_\infty$ and the 't Hooft Large N Expansion}
These examples suggest (correctly) that the general expansion has the form
\be 
\text{amplitude}= N^a \sum_{n=0}^{\infty}\frac{P_n(q)}{N^n} + NPC
\label{dkexpansion}
\ee
The value of $a$ depends on the particular amplitude. 
   $P_n(q)$ is an infinite order polynomial in $q,$ with the order of the first term increasing with $n$. 
   
   This is  analogous to the  large $N_{ym}$ expansion of gauge theories which has the form,
   \be  
\text{amplitude} = N^a\sum_n \frac{P_n(\alpha)}{{N_{ym}}^n} +NPC,
\label{hooft}
\ee
   where $\alpha$ is the 't Hooft coupling constant\footnote{Ordinarily the 't Hooft coupling constant is denoted $\lambda.$ Because $\lambda$  has already been used---for example in \eqref{ls3}---I have replaced it by $\alpha$ to denote the 't Hooft coupling. }.  By comparing  \eqref{dkexpansion} and \eqref{hooft} we see that 
 $q$ plays the  role of the  't Hooft coupling constant. This observation adds an interesting twist to the \dk \ formula  
$$ {q^2 } = \lambda  N.$$ It  parallels 't Hooft's definition
\be  
\alpha = g^2 N_{ym}.
\ee
One sees  that  $\lambda $ plays the role  of the gauge coupling $g^2$. 
Instead of the sum of planar diagrams the first term in \eqref{hooft} is the sum of all melon diagrams. 

Also note that the limit $\lambda = \text{fixed}$ is precisely analogous to the AdS/CFT limit  $g^2 = \text{fixed}$  i.e., the flat space limit.

\subsection{Relation to the 't Hooft and Flat-space Limits}

\begin{figure}[H]
\begin{center}
\includegraphics[scale=.7]{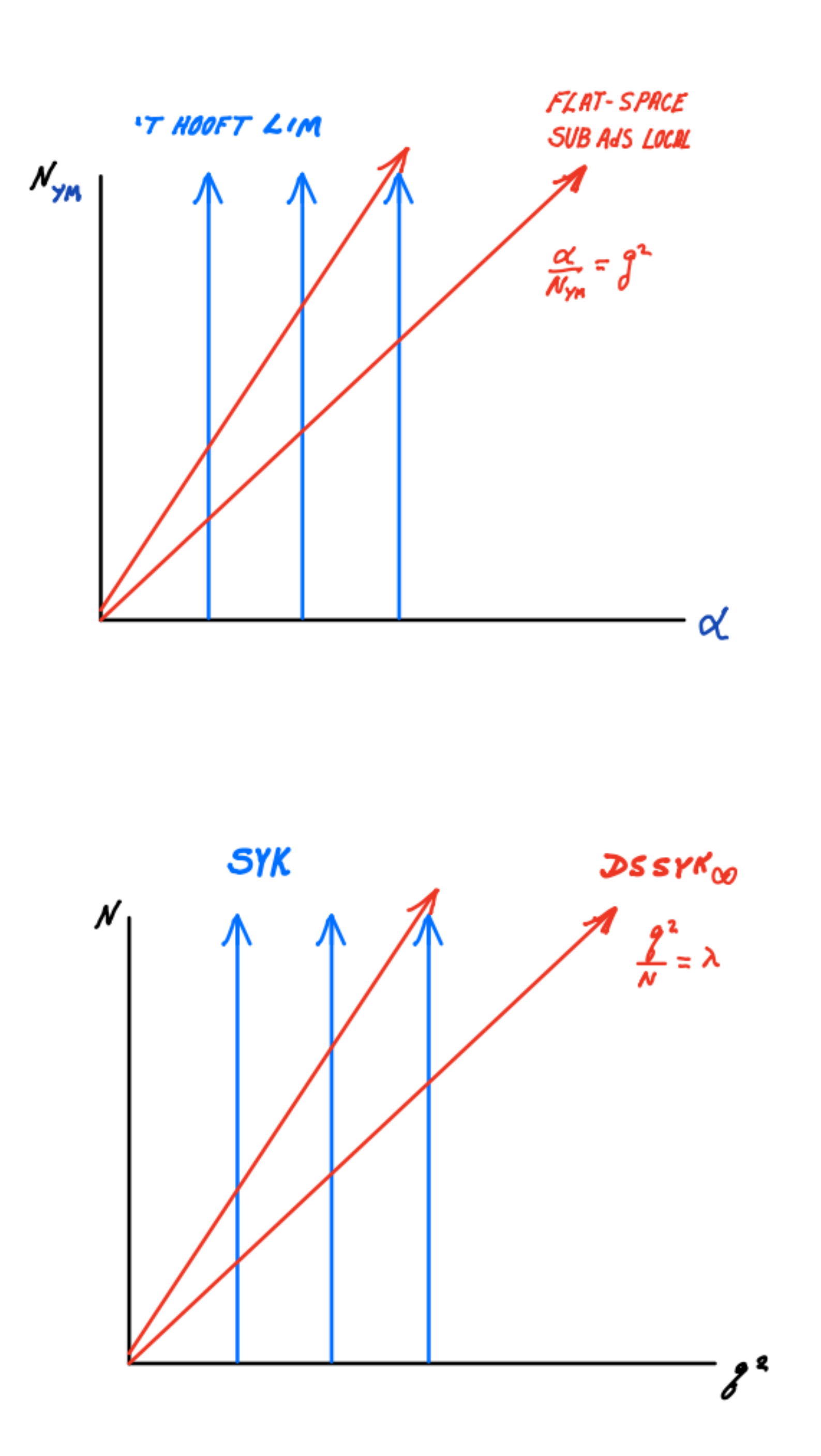}
\caption{The 't Hooft and flat-space limit of gauge theory parallel the fixed-$q$ and double-scaled limit of SYK.}
\label{limits}
\end{center}
\end{figure}

There are two types of limits that occur in both gauge theories and SYK theories. They are depicted in figure \ref{limits}.  In the upper panel referring to gauge theories the horizontal axis is the 't Hooft coupling and the vertical axis is $N_{ym}$. Moving along the vertical blue trajectories defines the conventional 't Hooft limit in which $N\to \infty$ while  $\alpha$ is held fixed. Almost all work on  large $N_{ym}$ gauge theory  is in this limit.
By considering large but fixed $\alpha$ one learns about the strongly coupled 't Hooft limit.

The flat-space limit is defined differently. It is the limit $N_{ym} \to \infty$ while holding fixed the gauge coupling constant $g.$ The flat-space limit is defined by moving along the red trajectories. It is notoriously difficult  to control and is much less studied than the 't Hooft limit.  It has some extraordinary counterintuitive properties \cite{Polchinski:1999yd}.
One feature of the flat-space limit in a holographic setting is that the theory becomes local on scales infinitely smaller than the AdS scale. That behavior is sometimes called ``sub-AdS locality."

In the bottom panel the same kind of diagram is shown for the SYK system at infinite temperature. The horizontal axis is $q^2$ and the vertical axis is $N$. The analog of the 't Hooft limit (moving vertically along blue trajectories) is the conventional fixed-$q$ limit. 

The double-scaled limit is defined by moving along the red trajectories in analogy with the flat-space limit of gauge theory. As might be expected it is a more difficult limit to explore.
 The conjecture of this paper is that the double-scaled limit is holographically dual to the semiclassical limit of JT de Sitter space. 

\sc
\subsection{The Emergent String Scale and Sub-dS Locality} \label{S:strscale}

In both  large-$N$ gauge theories and \dk  \  perturbation theory is IR divergent due to the bare masslessness of the fundmental constituents (the fermions in SYK, the gluons in gauge theory). However in both cases there is an emergent IR cutoff scale. 
In large-$N_{ym}$ QCD  the emergent scale is the confinement scale where the theory starts to behave with stringy characteristics. These include string-like   electric flux tubes, Hagedorn-like behavior, an energy gap,  Regge behavior, and confinement of non-singlets. None of this is
visible in  perturbation theory, but it could be seen if it were possible to  sum all planar diagrams.

To see that there is an emergent dynamical scale in \dk \ let us return to figure \ref{melon} and the infrared divergent expression \eqref{vacdiag2}. 
The diagram contains a numerical factor and an integral over the relative cosmic times of the two vertices.
 The bare propagators are $\epsilon(t_2-t_1)$ and when combined give an integrand which is independent of the relative time, thus leading to an IR divergence.
 
\begin{figure}[H]
\begin{center}
\includegraphics[scale=.4]{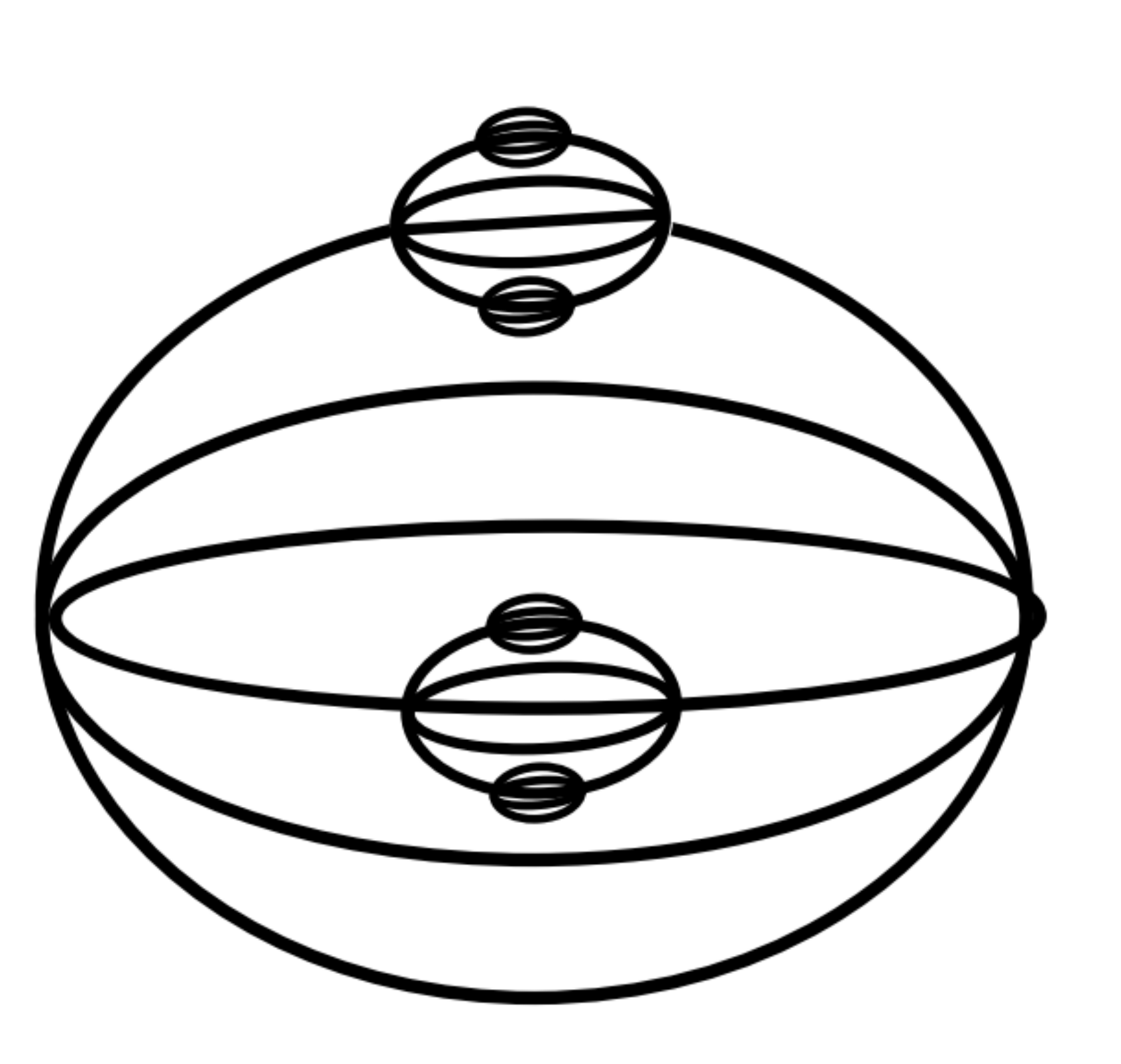}
\caption{An infinite class of diagrams that can be summed by solving the Schwinger-Dyson equation.}
\label{bubbles}
\end{center}
\end{figure}

 However, figure \ref{melon} is just the first of an infinite number of melonic diagrams  in which the propagators are corrected by additional melons,  melons within melons, ad infinitum as illustrated in figure \ref{bubbles}.
 Each of these diagrams is IR divergent but they can be formally summed  by solving the Schwinger-Dyson equations  \cite{Maldacena:2016hyu}. 
The result  is that the  trivial  integrand is replaced by the nontrivial function \cite{Roberts:2018mnp},
\be  
\frac{1}{\cosh^2{\CJ q |t_1-t_2|}} = \frac{1}{\cosh^2{\frac{|t_1-t_2|}{L_s} }}
\label{cosh2}
\ee
Two things are evident. The first is that  emergent mass and length scales appear,
\bea 
M_s &=&\CJ q \cr \cr
L_s \eq \frac{1}{\CJ q}
\label{Ls=1/Jq}
\eea

Secondly, the new scale regulates the infrared divergences. For example, the divergent integral 
in \eqref{vacdiag2} is replaced by,
\be 
\int dt \to \int dt  \frac{1}{\cosh^2{\CJ q t}} = \frac{1}{\CJ q}
\label{IRfinite}
\ee

Two points to note: First the correlation function depicted in figure \ref{bubbles}  and equation \eqref{cosh2}  is real with no oscillations. Evidently it represents an imaginary energy characteristic of overdamped systems. We will return to that in section  \ref{plasma}. The second point is that the figure represents the propagation of a cord, i.e, a system composed of $q$ elementary fermions.
\bn

I want to come to something which I think is very important.   
Consider the two definitions of $\lambda.$ On the gravitational side the definition of $\lambda$ is given by  $(L_m/L_s)^2$ (see \eqref{ls3}). On the \dk \ side it is given by $q^2/N.$  Comparing \eqref{Ls=1/Jq} and \eqref{Mm=JsqrtN} we see that the two definitions agree. To put it another way, the condition that the ratio of string  to micro scale be fixed and finite in the semiclassical limit is equivalent to the condition that $\lambda = q^2/N$ be fixed and finite in the limit $N\to \infty.$

If the string length does track the micro scale then one can be sure that it tends to zero in cosmic units. In other words the theory satisfies  the all-important requirement of  sub-dS locality. 

In AdS/CFT the situation is entirely analogous. In the 't Hooft limit of fixed $\alpha$ the string scale is of order the AdS scale, violating sub-AdS locality. But in the flat-space fixed $g$ limit the  
 the string scale is fixed in Planck units, thereby satisfying sub-AdS locality.

The bottom line is 
 that the double-scaled limit (for de Sitter space) is not an arbitrary assumption.   It  may be derived from the physical requirement of sub-ds locality.
 
 I think that is worth repeating:
 
 \bn
 \it
 The double-scaled limit, $\frac{q^2}{N} = \text{finite}$ for de Sitter space is not an arbitrary assumption. It follows from the physical requirement of sub-ds locality.
 \rm
 \bn

\subsection{Single Fermion Dressed Propagator}
The Diagram in figure \ref{bubbles} can be written as the $q$th power of the dressed fermion propagator. Equivalently the dressed propagator is given by,
\be 
 \la \chi_i(0) \chi_j(t)   \ra =\lf \frac{1}{\cosh^2{\CJ q |t_1-t_2|}} \rg^{\frac{1}{q}}
 \label{1overq}
\ee
In the large $q$ limit this becomes,
\be 
 \la \chi_i(0) \chi_j(t)   \ra = e^{-2\CJ t}
\label{prop}
 \ee
 Again note that this is real with no oscillations.

\subsection{Imaginary Masses}
In  equations \ref{1overq} and \ref{prop} we see  correlation functions that exponentially decay without any tendency to oscillate. That may seem  surprising. Ordinarily a correlation function will oscillate with a frequency characteristic of the energy carried by the intermediate state. In \ref{1overq} and \ref{prop} the correlation behaves as if the energy were pure imaginary.  

\begin{figure}[H]
\begin{center}
\includegraphics[scale=.5]{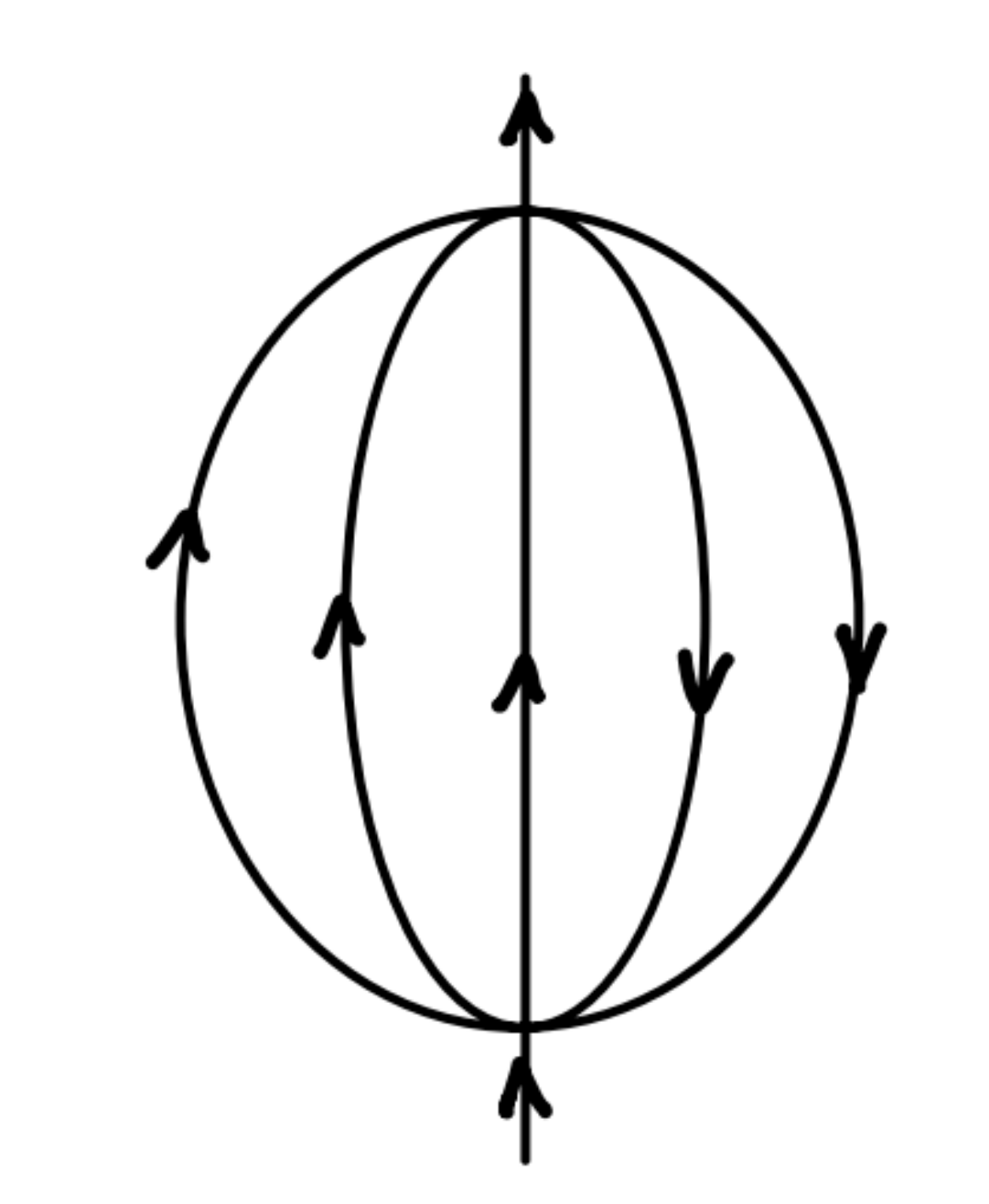}
\caption{Self-energy diagram for a fundamental fermion}
\label{firmass}
\end{center}
\end{figure}

For the case of the single fermion we can calculate the self-energy diagram in figure \ref{firmass} for the complex SYK model.
The diagram is given by
\be 
\gamma =2\frac{N^{q-1}}{(q-1)!}   \frac{N^q}{q!} \times  \CJ^2  \frac{q!}{N^q}  \frac{q!}{N^q}N \times  \frac{1}{q \CJ} 
\label{m=}
\ee
where the first factor is a combinitoric counting factor, the second factor is the propagator represented by the broken red line, and the third factor is the integral over time as in \eqref{IRfinite}. The result is
\be 
\gamma =2 \CJ
\label{gamma=}
\ee
consistent with \eqref{prop}.

\section{Confinement} \label{pair energy}

We will now turn to the question of confinement.

\subsection{Single Fermions}

Consider the correlation function between fermion operators at two times
$$ \la  \bx_i(0) \chi_i(t)   \ra .$$

\begin{figure}[H]
\begin{center}
 \includegraphics[scale=.5]{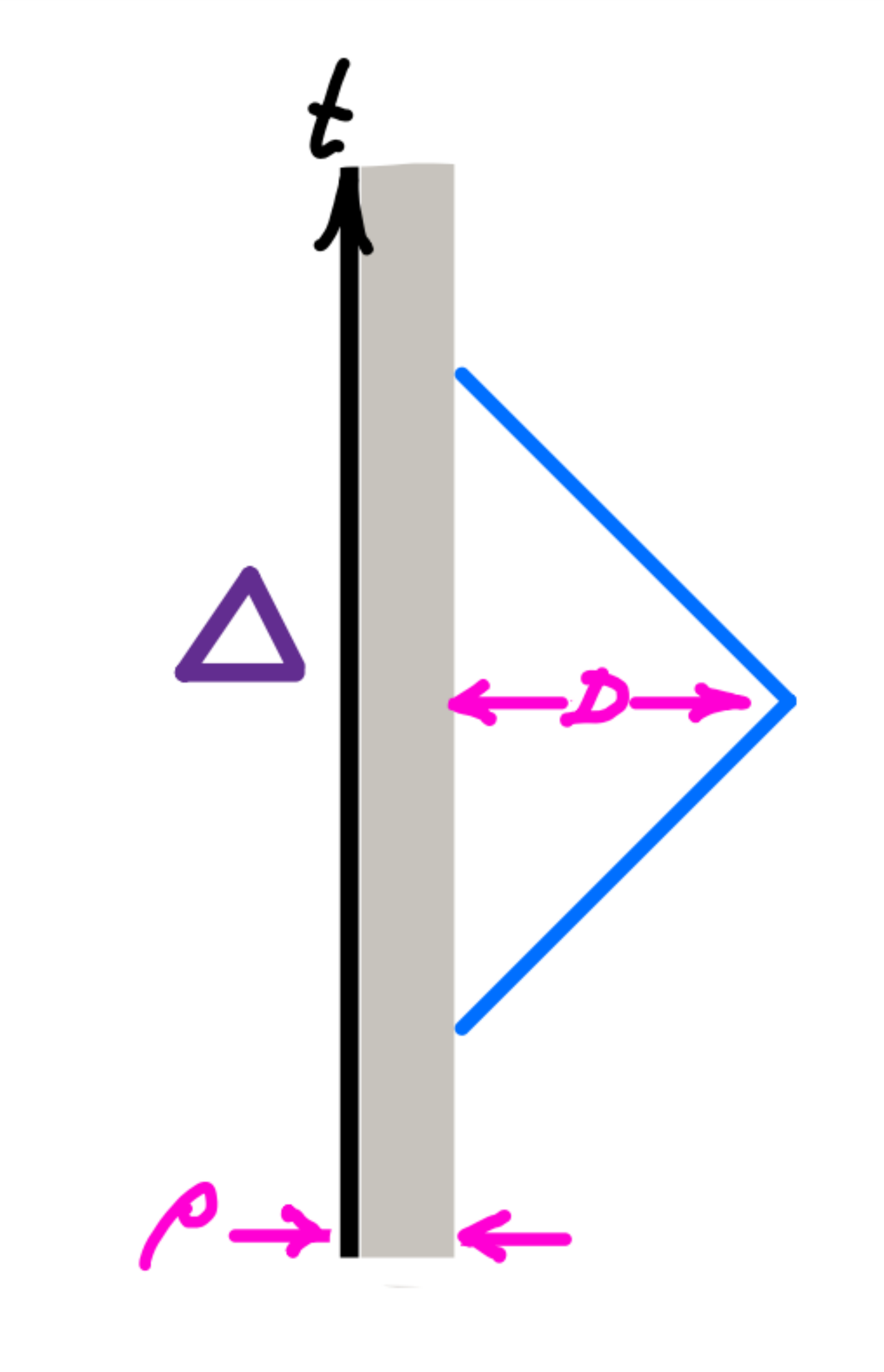}
\caption{The horizon, the stretched horizon and a fluctuation lasting time $\Delta$.} 
\label{blue}
\end{center}
\end{figure}

It has the form,
\be
 \la  \bx_i(0) \chi_i(t)   \ra  = e^{-\gamma t}.
\ee
We may think of it as the amplitude that a particle emitted from the stretched horizon is absorbed after a time $\Delta.$
The question is how deep into the bulk of the static patch can a fluctuation be felt. Clearly the answer is no deeper than the distance that the blue triangle in figure \ref{blue}, (bounded on the left by the stretched horizon and on the right by null rays)  extends into the bulk. In other words the the distance $D$. 

Using the de Sitter metric it is easy to calculate  $D$ in terms of  $\Delta .$  The answer is,
\be 
D=\rho e^{\frac{\Delta}{L_c}}
\ee
Since the correlation is negligibly small for $\Delta > 1/\gamma$, the fluctuation can only be felt to distance
\be 
D\leq \rho \ e^{1/(\gamma L_c)}.
\ee

Going back to \eqref{prop} we find that for a single fermion $\gamma = 2 \CJ.$
It follows that,
\be  
D = \rho e^{2 \CJ L_c } \approx \rho
\label{D=}
\ee
Allthough the time interval $\Delta$ is cosmic (of order $L_c = 1/\CJ$) the distance $D$ is  the microscopic  width of the stretched horizon. If we suppose that width is  the string scale, then the fermion is confined to within a string length of the stretched horizon. 
This is exactly what we want; the huge number of fermion species confined to the stretched horizon.

\subsection{Fermion Pairs in Complex SYK}

We will use a similar argument for $U(1)$-neutral  fermion pair operators.
Fermion pair operators  can be classified by their transformation properties under 
$SU(N)$---the symmetry of ensemble averages. The single fermion itself transforms under the fundamental. Fermion pairs can transform as adjoints,
\be 
\text{adjoint} = A = \sum_{ij}  T_{ij}    \bx_i \chi_j
\label{adj}
\ee
(where $T_{ij}$ is a traceless matrix) or as singlets 
 of $SU(N)$,
\be 
\text{singlet} = S = \frac{1}{\sqrt{N}} \sum_{ii}     \bx_i \chi_i
\label{singlet}
\ee

Almost all fermion pairs are adjoints. The number of orthogonal adjoints is $(N^2 -1)$  whereas the singlet consists of only a single operator.

\begin{figure}[H]
\begin{center}
\includegraphics[scale=.3]{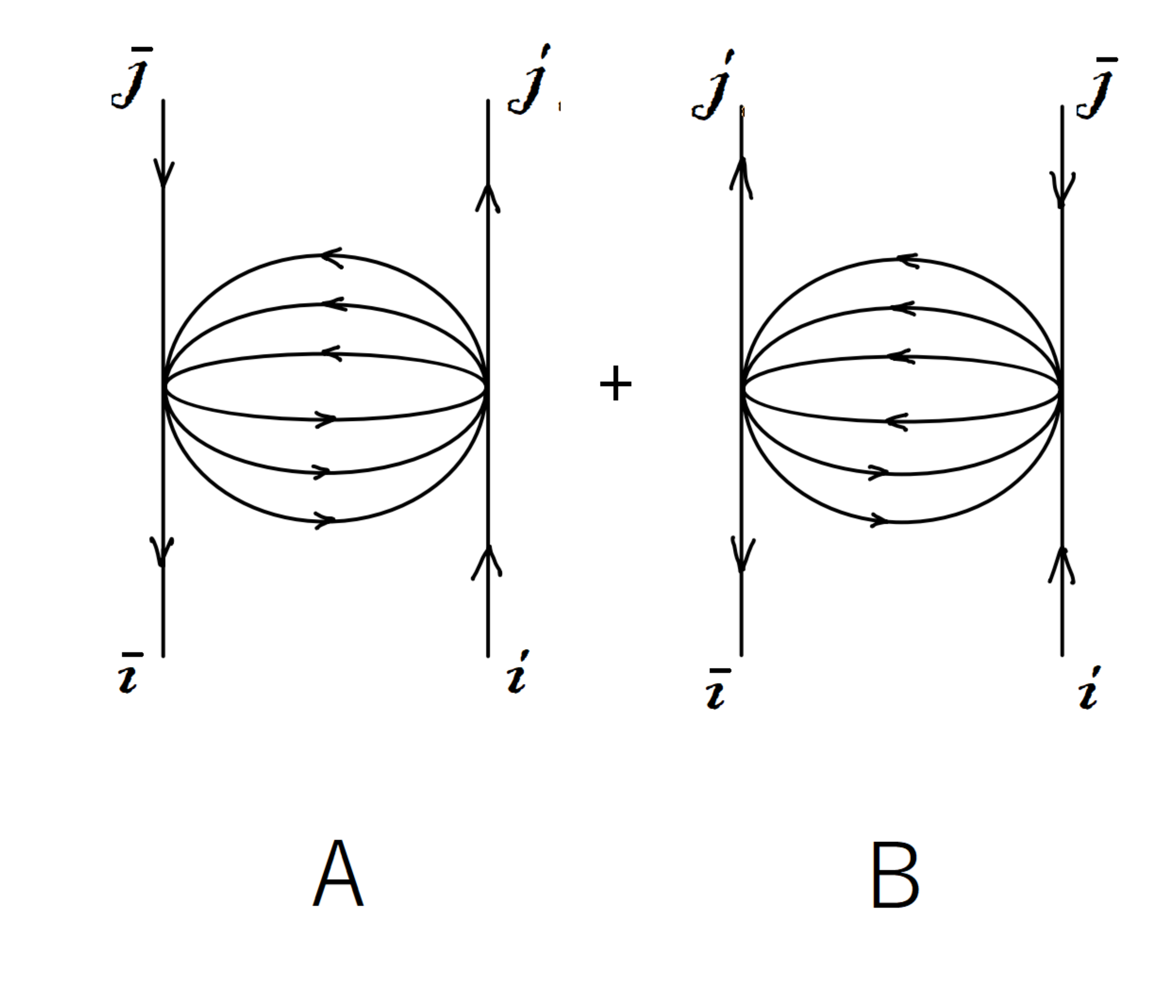}
\caption{Leading diagrams for the interaction of a fermion pair.}
\label{pairdiag}
\end{center}
\end{figure}

 The leading diagrams for the interaction of a pair are shown in figure \ref{pairdiag}.

We will think of the space of operators $\bx_i \chi_j$ as a vector space $|i,j).$ There is a subspace of this vector space spanned by ``diagonal" operators of the form $|i,i) = |i).$  Note that the transitions  
 described by diagrams A and B stay within this diagonal subspace.
 
 The diagonal subspace contains $(N-1)$ adjoint operators,
 \be  
 |\text{adjoint}) = \sum T_i |i) \ \ \ \ \ \ \ \ \ \  \sum_i T_i =0
 \ee
 and a singlet,
 \be 
 |\text{singlet}) = \sum_i |i)
 \ee
The processes in diagram  A allow  transitions from any $|i)$ to any $|j)$. The  matrix describing them has the form,
\be 
\sum_{ij} |i) (j|
\ee
This is proportional to a projection operator $\Pi_{singlet}$ onto the singlet state.
\be 
\sum_{ij} |i) (j| = N \  \Pi_{singlet}
\ee

Diagram B allows transitions from $|i)$ to $|j)$ as long as $i \neq j.$ The matrix describing such transitions has the form,
\be 
\sum_{i\neq j} |i) (j| = N  \Pi_{singlet} - I
\ee
where $I$ is the identity matrix.

We can therefore write diagram A in the form,
\be 
\text{Diag A} = 4 \ \frac{N^{q-1}}{(q-1)!}  \  \frac{N^{q-1}}{(q-1)!} \times \CJ^2 N \frac{q!}{N^q}  \frac{q!}{N^q}   \times    \frac{1}{q\CJ} \times N \ \Pi_{singlet}
\ee
where $ \Pi_{singlet}$ is the projection matrix onto the singlet pair $\sum_i \bx_i \chi_i.$

Diagram B has the opposite sign and can be written,
\be 
\text{Diag B} = 4 \ \frac{N^{q-2}}{(q-2)!}  \  \frac{N^{q}}{q!} \times \CJ^2 N \frac{q!}{N^q}  \frac{q!}{N^q}   \times    \frac{1}{q\CJ} \times (N \ \Pi_{singlet} -I  )
\ee

The leading terms of A and B cancel\footnote{I am grateful to Juan Maldacena for pointing this out.} leaving,
\be 
4\lf   \CJ \Pi_{singlet} + (q-1)\frac{\CJ}{N} I                \rg
\ee
In the limit $N\to \infty, \ \ q^2 \sim N$ the second term goes to zero. Thus the final result is 

\be 
\text{A+B} = 4\CJ \ \Pi_{singlet}
\ee

However this is not the whole story. There is one more term representing the single particle (imaginary) energies in \eqref{gamma=}. These give $4\CJ I.$ Thus the full energy matrix is, 
\be 
\Gamma =4\CJ I -4 \CJ \  \Pi_{singlet} .
\ee

From this we read off,
\bea  
\gamma_{adjoint} \eq 4\CJ \cr \cr
\gamma_{singlet} \eq 0.
\eea

One may wonder what happens to the off-diagonal pairs like $\bx_1 \chi_2$. These all transform as adjoints and by the $SU(N)$ symmetry of the ensemble average they have the same energy as the diagonal adjoints.

The meaning of this is clear: The adjoints (all $N^2 - 1$ of them) have imaginary energy $4\CJ$ and by the argument accompanying figure \ref{blue} are confined.
The singlet however has energy zero and is therefore not confined.

Naively one might expect the singlet correlation to also be given by $e^{-4\CJ t}$
but that is not possible. The reason is that the singlet operator happens to be the total $U(1)$ charge. It follows that the singlet-singlet correlation must be time-independent,
\be 
\frac{d}{dt} \la S(0) S(t) \ra = 0.
\label{constant}
\ee
It can be easily calculated at $t=0$ and one finds that it's value is exactly $1$. Thus for all time,
\be 
 \la S(0) S(t) \ra = 1.
 \label{trivial}
\ee
However, unlike the other correlation functions---single fermions and adjoints---the singlet correlation has no imaginary part. This means that it does not represent a causal propagator. The trivial behavior of the singlet is closely connected to the fact that in $(1+1)$-dimensions the massless photon is non-dynamical

The photon can be made dynamical by giving it a small mass which can be done by breaking the $U(1) $ symmetry\footnote{This is now being studied by myself and Adel Rahman. }. There are many ways to do this. One example would be to introduce the symmetry breaking term,
\be 
H_{breaking} = \sum K_{i_1, i_2....i_{2q}} \chi_{i_1}\chi_{i_2}.....\chi_{i_{2q}} + \text{hc}
\ee
By making $K$ small enough the mass of the photon can be kept less than $\CJ$ which is essential if the photon is to be radiated as Hawking radiation.

\section{Analogy with QCD Plasma} \label{plasma}

A pattern emerges from what we've seen above. Both the single fermion and the adjoint pair are confined to the vicinity of the stretched horizon. This leads to the question: What is the organizing principle which determines which combination of fermions are confined? I will propose an answer based on symmetry.

\subsection{SU(N) Confinement}

Both single fermions and adjoints are confined, and they  are  also non-singlets under the $SU(N)$ symmetry of the ensemble average. The singlet by contrast is not confined and may be emitted into the bulk.

As I've explained the states in \dk \ can be organized into $SU(N)$ multiplets and in the ensemble-averaged theory the properties of the states are degenerate within a multiplet. For example in the  ensemble averaged theory all $N$ single fermion states have the same mass. Within a single instance of the ensemble this is not true, but the splittings within a multiplet are expected to be $\frac{1}{N}$ effects\footnote{I thank Henry Lin for a discussion on this point.}.

The same is true for the $(N^2-1)$ adjoint states. Likewise if one member of a multiplet is confined (unconfined), all members of that multiplet are confined (unconfined). I therefore propose the principle:

\bn
\it  Only $SU(N)$ singlets are unconfined and can propagate freely into the bulk of the static patch. All multiplets with non-trivial $SU(N)$ charge are confined. 
\rm

\bn

Given the sparsity of singlets, this would solve the problem of an overabundance of degrees of freedom in the  bulk of the static patch.

\subsection{QCD Plasma Analogy}\label{plasma}

\begin{figure}[H]
\begin{center}
\includegraphics[scale=.3]{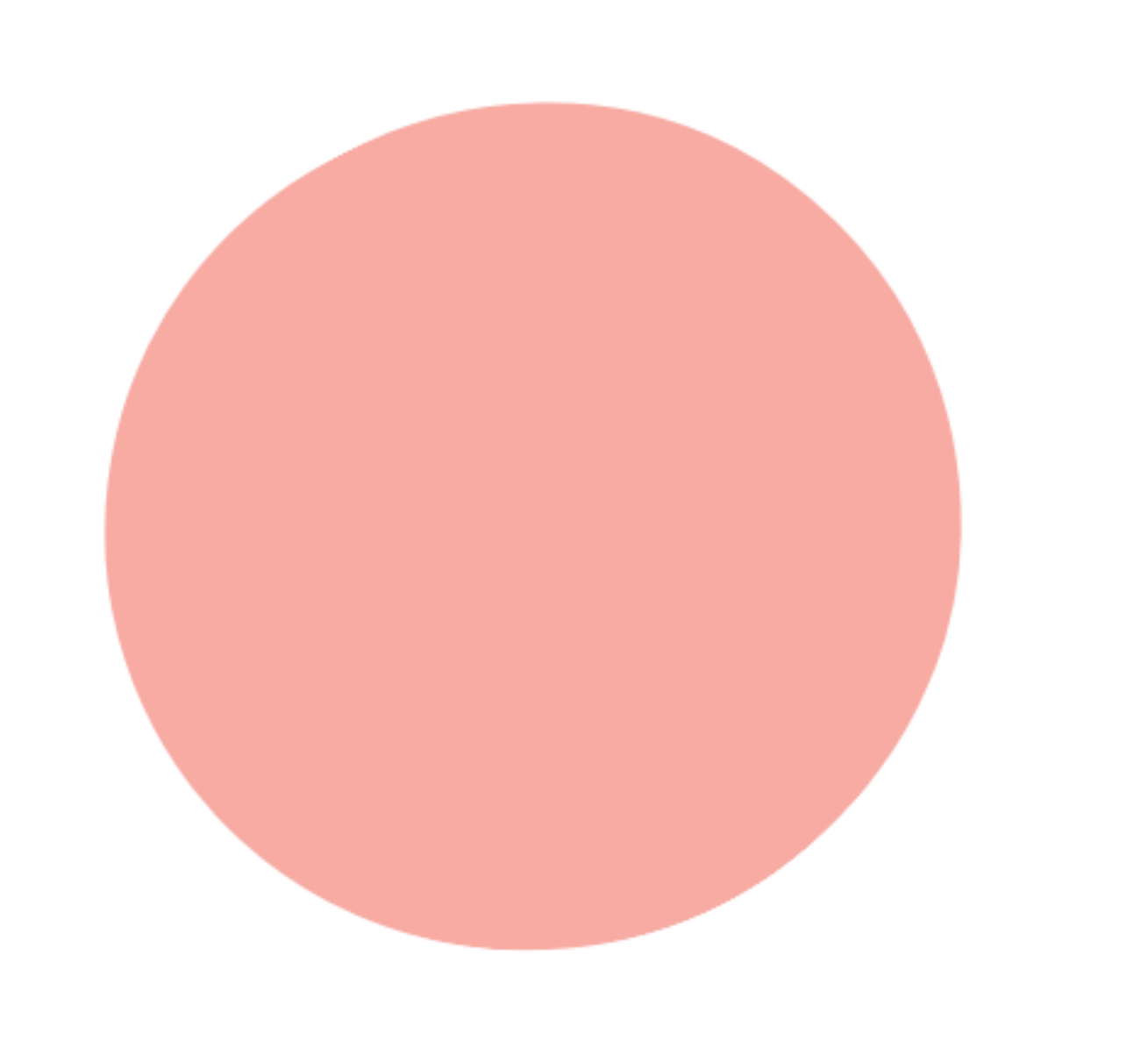}
\caption{A bubble of QCD plasma}
\label{plasma1}
\end{center}
\end{figure}

\begin{figure}[H]
\begin{center}
\includegraphics[scale=.3]{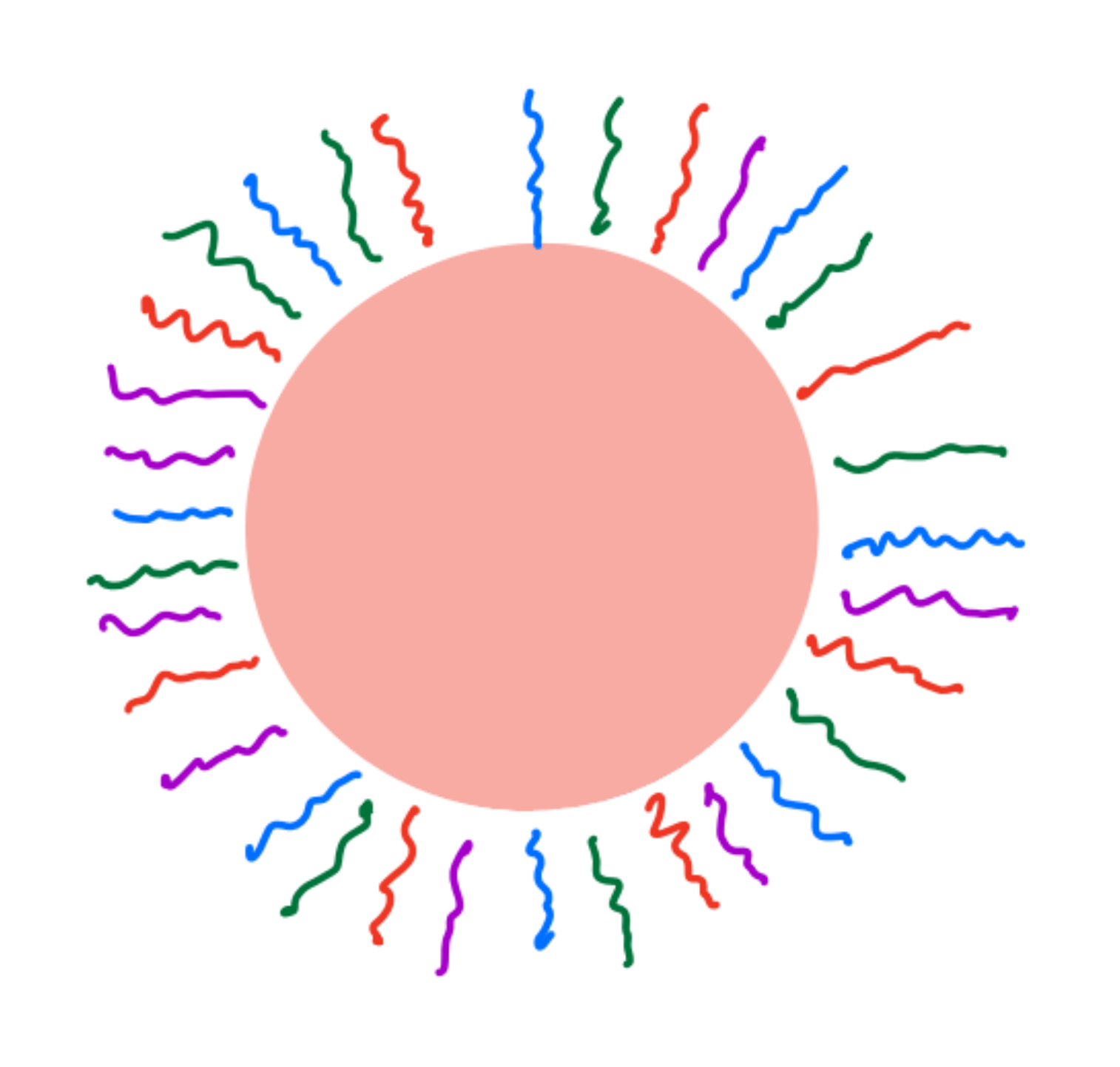}
\caption{Plasma bubble radiating gluons. The luminosity is of order $N_{ym}^2$}
\label{plasma2}
\end{center}
\end{figure}

There is a close analogy with quark confinement in large-N QCD.
Consider a  bounded region of space in which there is bubble of large-N QCD plasma as in figure \ref{plasma1}.
One might question whether such a bubble can be stable with respect to very rapid evaporation into gluons,  given that  the number of gluon species, and therefore the luminosity, is  $N_{qcd}^2$ (see figure \ref{plasma2}).

\begin{figure}[H]
\begin{center}
\includegraphics[scale=.3]{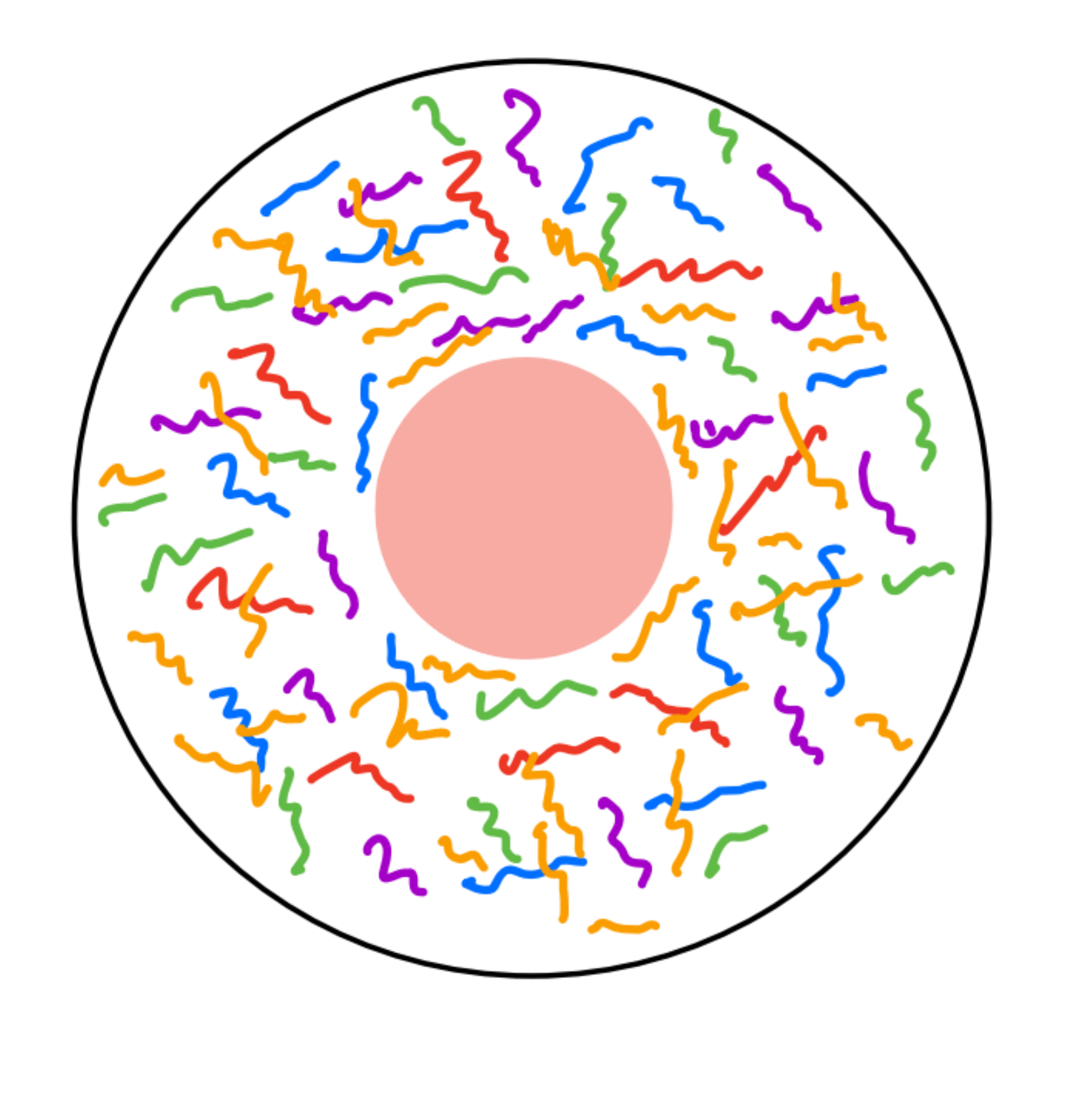}
\caption{Atmosphere between bubble and box.}
\label{plasma3}
\end{center}
\end{figure}

\begin{figure}[H]
\begin{center}
\includegraphics[scale=.3]{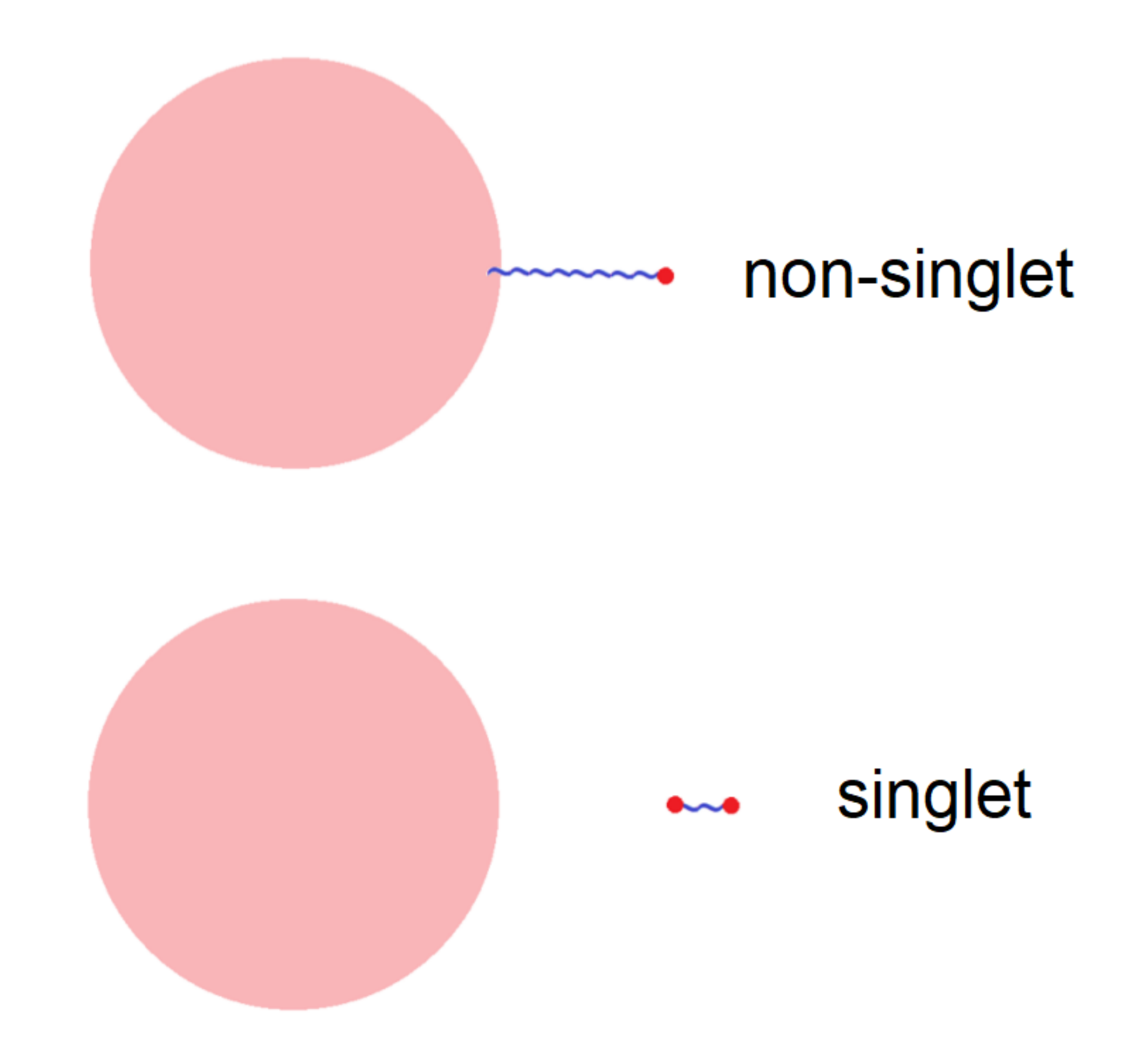}
\caption{A quark is attached to a bubble of QCD plasma by an electric string. A gauge-singlet such as a meson or glueball is free to evaporate from the bubble.}
\label{plasma4}
\end{center}
\end{figure}

\bn
If the bubble were placed in a box with reflecting walls, as in figure \ref{plasma3} the atmosphere between the bubble and the walls of the box would be very dense
with $N_{ym}^2$ species in equilibrium with the hot bubble.
\bn
 But this not right: non-singlets are attached to the plasma by chromo-electric strings and are therefore confined. Only singlets like glueballs or mesons can be emitted (see figure \ref{plasma4})  and these are very sparse.  The number of species of these objects is order unity and does not grow with $N$. 
\bn
Therefore in the large $N$ limit the bubble  evaporates slowly.

For the same reason, if the bubble were contained in a larger box with reflecting walls, the atmosphere in the region between the bubble and the walls would be sparsely occupied by singlets.

The model of a gauge-theory plasma for the horizon \dof \ may be more than an analogy; it may be relevant to four-dimensional de Sitter space where the holographic theory has the form of a large-$N$ matrix quantum mechanics with a conventional $SU(N)$ symmetry   \cite{Banks:2006rx}\cite{Susskind:2021dfc}. Once again there are many more \dof \ than propagating modes in the bulk and most of them must be restricted to the horizon region. 
In this case there  are no random couplings or ensemble averages but the exact internal $SU(N)$ symmetry can provide confining forces which lift the energies of non-singlets, leaving only the singlets to propagate into the bulk. 

The QCD example does exactly what we want. The large $N$ degrees of freedom contribute to the entropy of the plasma bubble but not to the luminosity or the excessive number of particles in the thermal atmosphere.

This of course is not an explanation of why the ensemble symmetry of \dk \ provides confining forces but it is a useful analogy.

\subsection{Cords}

The focus of the previous section was on systems of small size; size-one for single fermions and size-two for pairs. Naively the energy of such systems is $\sim \CJ$ which makes them candidates for Hawking radiation. But as we have seen, almost all of these states are confined.

By contrast, recent work on the double-scaled limit \cite{Berkooz:2018jqr}\cite{Lin:2022rbf} has involved a different range of sizes and energies; namely the string scale. The systems of interest have size $\sim q$
and (naively)  energy $\sim q\CJ.$ They come in two kinds: Hamiltonian chords and matter cords. 

We wish to know whether cords are in any sense like particles which can be emitted into the bulk? In this respect there are two puzzling things about cords which seems not to have been addressed in the literature.

\begin{enumerate}
\item One expects cords to have an energy of order $q\CJ.$ Correlation functions of cord operators should therefore oscillate like,
\be 
\la C(0) C(t) \ra \sim e^{\pm i q\CJ t} .
\label{osc?}
\ee
But they don't. A typical correlation function exponentially decays but doesn't oscillate. They behave like overdamped oscillators. Figure \ref{bubbles} is an example of a cord correlation, and from
 \eqref{cosh2} it decays but does not oscillate.

\item One can also see from figure \ref{bubbles} that correlation functions factorize. The cord-cord correlation functions simply factorizes into products of single fermion correlation functions. For  operators of size $\Delta q$ the two point function has the form 
\be  
\la C_{\Delta}(0) C_{\Delta}(t) = \la \bx(0) \chi(t)\ra^{q \Delta }
\label{cordcord}
\ee
\end{enumerate}

We can summarize the behavior of cords as follows: 
First of all they are collections of almost non-interacting fermions whose correlation functions
 factorize into single fermion correlations.

Second, the fermions of which the cords are composed decay on a short time scale before they oscillate. They behave like over-damped oscillators.

These two properties may seem odd but they are easily understood in the QCD plasma analogy. First, in a dense hot plasma gauge forces are screened by the Debye effect. Even if the colored particles are confined in vacuum, in the plasma the forces between them can be neglected. Correlation functions factorize to a good approximation.

Second, in a plasma charged particles decay into the plasma. If the plasma is dense enough the charged fields will be over-damped. 
It therefore seems that cords as currently understood (as well as single fermions) behave as if they were propagating in a dense hot plasma. The plasma is located in the stretched horizon. Near the horizon the fermions propagate freely but when they try to escape the near horizon region they find themselves confined unless they form singlet configurations. That seems to be the physical picture of what a cord is in \dk.

A cord is a collection of $\sim \sqrt{N}$ fermions. Most cords will not be singlets but a small number may be. There is a QCD analogy. The baryon in large $N$ QCD is a collection of $N_{qcd}$ quarks. Keeping in mind that $N_{qcd}$ is analogous to $\sqrt{N}$ in SYK, we see that the baryon is similar to a singlet cord. Baryons can escape and propagate outside the plasma bubble. It seems reasonable to suppose that singlet cords can propagate into the static patch. But that is a tiny fraction of all multi-fermion states.

\section{Summary}\label{summary}

In this concluding section I will summarize  the important points of this paper. The first and most important is the fundamental conjecture itself, namely that \dk \ is a holographic description of de Sitter space.
But let me be  more fine-grained.

\begin{enumerate}
\item  In the semiclassical limit of de Sitter space there is a separation of scales into cosmic, string, micro, and maximum scales. Precisely the same separation occurs in \dk.  The correspondence is described in the dictionary in section \ref{dictionary}.
\item Working in cosmic units, the perturbation expansion of SYK has the same form as the 't Hooft large-$N$ expansion of non-abelian gauge theory. The two interesting limits of gauge theory---the 't Hooft limit and the flat space limit---are paralleled by the fixed $q$ limit of SYK, and the double scaled limit of \dk. 
\item There is a need for a confinement mechanism to eliminate the large number of species that might otherwise propagate into the bulk of the dS static patch.  The string scale  of large $q$ SYK is the analog of the QCD confinement scale.
\item A point that I've emphasized in section \ref{S:strscale} is that the double scaled limit is a natural consequence of requiring a sensible local bulk theory in which the string length scale is parametrically microscopic compared to the cosmic scale. The double scaled limit can be derived from the assumption of sub-dS locality.
\item In section \ref{pair energy} evidence was presented for the basic confinement conjecture; namely that $SU(N)$ charge is confined but $SU(N) $ singlets are not confined. The evidence for the latter is not as strong as it could be. I suggested that breaking the $U(1)$ symmetry of complex SYK would allow  the photon to become dynamical. It this context a sharp test will be possible if correlation functions of the singlet can be computed for times of order $L_c \log{N}.$
\item  In section \ref{plasma} I explained that fermions and cords behave like  particles in a hot dense plasma. That  explains  both the  factorization of correlation functions and their overdamped behavior.

This last point while counterintuitive in \dk \ is natural in the gravitational theory. The proper Unruh temperature at the stretched horizon is $\sim 1/\rho$ where $\rho $ is the microscopic thickness of the stretched horizon. From that point of view the stretched horizon should behave like a hot medium. What is surprising is this property shows up clearly in \dk.

\end{enumerate}

There are two questions that I didn't address. The first is: what is the geometry behind the horizon? If \dk \ is a description of de Sitter space then the geometry behind the horizon should exponentially expand. The only way that I know of to approach this question is through the growth of complexity. In
\cite{Susskind:2021esx} I explained that the exponential growth translates into ``hyperfast" complexity growth. It is important to verify that the complexity growth in \dk \ is hyperfast. 

The second unaddressed question concerns the interpolation between low and infinite temperatures. At low temperatrure the geometry of large $q$ SYK is thought to be a long $AdS(2)$ throat with negative curvature. If the geometry at infinite temperature is de Sitter then the question is what lies in between? 

I hope to return to both of these issues in the near future.

\section*{Acknowledgements}
I thank Henry Lin,   Adel Rahman,   and Edward Witten for  discussions. I especially want to thank Juan Maldacena for pointing out an error in an earlier version of this paper.


\begin{thebibliography}{99} 
\bibitem{Susskind:2022bia}
L.~Susskind,
``De Sitter Space, Double-Scaled SYK, and the Separation of Scales in the Semiclassical Limit,''
[arXiv:2209.09999 [hep-th]].

\bibitem{Rahman:2022jsf}
A.~A.~Rahman,
``dS JT Gravity and Double-Scaled SYK,''
[arXiv:2209.09997 [hep-th]].

\bibitem{Berkooz:2018jqr}
M.~Berkooz, M.~Isachenkov, V.~Narovlansky and G.~Torrents,
``Towards a full solution of the large N double-scaled SYK model,''
JHEP \textbf{03}, 079 (2019)
doi:10.1007/JHEP03(2019)079
[arXiv:1811.02584 [hep-th]].

\bibitem{Lin:2022rbf}
H.~W.~Lin,
``The bulk Hilbert space of double scaled SYK,''
JHEP \textbf{11}, 060 (2022)
doi:10.1007/JHEP11(2022)060
[arXiv:2208.07032 [hep-th]].

\bibitem{Maldacena:2016hyu}
J.~Maldacena and D.~Stanford,
``Remarks on the Sachdev-Ye-Kitaev model,''
Phys. Rev. D \textbf{94}, no.10, 106002 (2016)
doi:10.1103/PhysRevD.94.106002
[arXiv:1604.07818 [hep-th]].

\bibitem{Cotler:2016fpe}
J.~S.~Cotler, G.~Gur-Ari, M.~Hanada, J.~Polchinski, P.~Saad, S.~H.~Shenker, D.~Stanford, A.~Streicher and M.~Tezuka,
``Black Holes and Random Matrices,''
JHEP \textbf{05}, 118 (2017)
[erratum: JHEP \textbf{09}, 002 (2018)]
doi:10.1007/JHEP05(2017)118
[arXiv:1611.04650 [hep-th]].

\bibitem{Dong:2018cuv}
X.~Dong, E.~Silverstein and G.~Torroba,
``De Sitter Holography and Entanglement Entropy,''
JHEP \textbf{07}, 050 (2018)
doi:10.1007/JHEP07(2018)050
[arXiv:1804.08623 [hep-th]].

\bibitem{Chandrasekaran:2022cip}
V.~Chandrasekaran, R.~Longo, G.~Penington and E.~Witten,
``An algebra of observables for de Sitter space,''
JHEP \textbf{02}, 082 (2023)
doi:10.1007/JHEP02(2023)082
[arXiv:2206.10780 [hep-th]].

\bibitem{Lin:2022nss}
H.~Lin and L.~Susskind,
``Infinite Temperature's Not So Hot,''
[arXiv:2206.01083 [hep-th]].

\bibitem{Berkooz:2020uly}
M.~Berkooz, V.~Narovlansky and H.~Raj,
``Complex Sachdev-Ye-Kitaev model in the double scaling limit,''
JHEP \textbf{02}, 113 (2021)
doi:10.1007/JHEP02(2021)113
[arXiv:2006.13983 [hep-th]].

\bibitem{Susskind:2020fiu}
L.~Susskind,
``Electric Forces in the Charged SYK Model,''
[arXiv:2012.12326 [hep-th]].

\bibitem{Susskind:2022dfz}
L.~Susskind,
``Scrambling in Double-Scaled SYK and De Sitter Space,''
[arXiv:2205.00315 [hep-th]].

\bibitem{Roberts:2018mnp}
D.~A.~Roberts, D.~Stanford and A.~Streicher,
``Operator growth in the SYK model,''
JHEP \textbf{06}, 122 (2018)
doi:10.1007/JHEP06(2018)122
[arXiv:1802.02633 [hep-th]].

\bibitem{Polchinski:1999yd}
J.~Polchinski, L.~Susskind and N.~Toumbas,
``Negative energy, superluminosity and holography,''
Phys. Rev. D \textbf{60}, 084006 (1999)
doi:10.1103/PhysRevD.60.084006
[arXiv:hep-th/9903228 [hep-th]].

\bibitem{Banks:2006rx}
T.~Banks, B.~Fiol and A.~Morisse,
``Towards a quantum theory of de Sitter space,''
JHEP \textbf{12}, 004 (2006)
doi:10.1088/1126-6708/2006/12/004
[arXiv:hep-th/0609062 [hep-th]].

\bibitem{Susskind:2021dfc}
L.~Susskind,
``Black Holes Hint Towards De Sitter-Matrix Theory,''
[arXiv:2109.01322 [hep-th]].

\bibitem{Susskind:2021esx}
L.~Susskind,
``Entanglement and Chaos in De Sitter Space Holography: An SYK Example,''
JHAP \textbf{1}, no.1, 1-22 (2021)
doi:10.22128/jhap.2021.455.1005
[arXiv:2109.14104 [hep-th]].

\end{thebibliography}
\end{document}